\documentstyle{article}
\newcommand{\mb}[1]{\mbox{\normalsize\boldmath $#1$}}
\newcommand{\One}{\hbox{\bf 1}}
\newcommand{\I}{\mb{I}}
\newcommand{\Frac}[2]{\leavevmode\kern.1em\raise.5ex
\hbox{\the\scriptfont0 #1}
\kern-.1em/\kern-.15em\lower.25ex\hbox{\the\scriptfont0 #2}}

\def\diag{\mathop{\rm diag}}
\def\Ord{{\cal O}}
\def\Lag{{\cal L}}
\def\SU{{\rm SU}}

\makeatletter
\newcounter{alphaequation}[equation]
 \def\thealphaequation{\theequation\hbox to
0.6em{\hfil\alph{alphaequation}\hfil}}
\def\eqnsystem#1{
\def\@eqnnum{{\rm (\thealphaequation)}}
\def\@@eqncr{\let\@tempa\relax
\ifcase\@eqcnt \def\@tempa{& & &}
\or \def\@tempa{& &}\or \def\@tempa{&}\fi\@tempa
\if@eqnsw\@eqnnum\refstepcounter{alphaequation}\fi
\global\@eqnswtrue\global\@eqcnt=0\cr}
\refstepcounter{equation}
\let\@currentlabel\theequation
\def\@tempb{#1}
\ifx\@tempb\empty\else\label{#1}\fi
\refstepcounter{alphaequation}
\let\@currentlabel\thealphaequation
\global\@eqnswtrue\global\@eqcnt=0
\tabskip\@centering\let\\=\@eqncr
$$\halign to \displaywidth\bgroup
  \@eqnsel\hskip\@centering
  $\displaystyle\tabskip\z@{##}$&\global\@eqcnt\@ne
  \hskip2\arraycolsep\hfil${##}$\hfil&
  \global\@eqcnt\tw@\hskip2\arraycolsep
  $\displaystyle\tabskip\z@{##}$\hfil
  \tabskip\@centering&\llap{##}\tabskip\z@\cr}

\def\endeqnsystem{\@@eqncr\egroup$$\global\@ignoretrue}
\makeatother

\begin{document}\begin{titlepage}
\large\hfill\vbox{\baselineskip12pt
 \hbox{\bf IFUP -- TH 72/94}
 \hbox{\bf UCB-PTH-94/29}
 \hbox{\bf hep-ph/9501334}
 \hbox{\bf LBL 36381}
 \hbox{January 1995}}
\vspace{7mm}
\begin{center}\vglue 0.6cm{\huge\bf\vglue 10pt
Violations of lepton flavour and CP\\
in supersymmetric unified theories\footnote{Supported in part by
U. S. Department of Energy under
Contract DE-AC03-76SF00098 and in part by the National Science
Foundation under grant PHY-90-21139.}\\}
\vglue 1.5cm
{\large\bf  Riccardo Barbieri$^\dagger$, Lawrence Hall$^\ddagger$ \rm and
       \bf Alessandro Strumia$^\dagger$\\}
\vglue 0.8cm
$\dagger$~{\em Dipartimento di Fisica, Universit\`a di Pisa\\}
{\rm and \em INFN, Sezione di Pisa, I-56126 Pisa, Italy\\}
\vglue 0.4cm
$\ddagger$~{\em Department of Physics,
University of California at Berkeley, California 94720}

\vfill
{\large\bf Abstract}
\end{center}
\vglue 0.3cm{\rightskip=3pc \leftskip=3pc \tenrm\baselineskip=12pt
\noindent\large As a consequence of the large top quark Yukawa coupling,
supersymmetric unified theories with soft supersymmetry breaking terms
generated at the Planck scale predict lepton flavour and CP violating
processes with significant rates.\\ \indent
The flavour violating
parameters of the low energy theory are derived in both~SU(5) and~SO(10)
theories, and are used to calculate the rate for $\mu\to e\gamma$.
The sensitivity of the search for $\mu\to e\gamma$ is compared with that
for $\mu\to e$ conversion in atoms, $\tau\to\mu\gamma$ and
the electric dipole moment of the electron.
The experimental search for these processes is shown to provide
a very significant test of supersymmetric unification,
especially in~SO(10) but also in~SU(5).}
\vfill\vfill
\end{titlepage}

\section{Introduction}
The importance of looking for direct tests of unified theories
cannot be overestated.
As is well known, such an opportunity is essentially restricted
to the study of violations of those conservation laws which are
valid in the Standard Model as a consequence of exact ``accidental''
global symmetries.
We refer to baryon number, $B$, and to the individual lepton numbers,
$L_e$, $L_\mu$ and $L_\tau$.

In these respects, the violation of individual
lepton numbers while preserving
the overall lepton number, $L=L_e+L_\mu + L_\tau$,
--- hereafter called Lepton Flavour Violation (LFV) ---
plays a special role.
If the Grand Unified Theory,
characterized by a large mass scale $M_{\rm G}$, has
the pure Standard Model as its low energy approximation, the rates
for the corresponding LFV processes
($\mu\to e\gamma$, $\mu\to e$ conversion, $\mu\to 3e$, etc.)
are unobservably small, since they are necessarily mediated by
non-renormalizable effective interactions scaled by inverse
powers of~$M_{\rm G}$.
On the contrary, in a supersymmetric unified theory
with supersymmetry effectively broken at the Fermi scale,
$m=\Ord(G_{\rm F}^{-1/2})$,
the rates for the LFV low energy processes
are only suppressed by powers of $1/m$ \cite{HKR}.
In general this would actually also be the case for
$B$ and/or $L$-violating processes, like proton decay, strongly
suggesting the need of matter parity (or $R$-parity)
in a unified supersymmetric theory.
Correspondingly, the LFV processes, consistent with matter parity
unlike $B$ and/or $L$ violations,
emerge as very interesting possible experimental signals
of supersymmetric unification.

In a previous paper \cite{BH}, two of us (R.B.\ and L.H.) have pointed out
that the large Yukawa coupling of the top quark at the unification scale,
$\lambda_{t\rm G}$, is an important source of flavour violation which
reflects itself, via the unified couplings, in relatively large rates for
general LFV processes.
There it has been argued that the study of the corresponding experimental
signals provides a test of supersymmetric unification at least as
significant as the one that can be obtained from either proton decay or
neutrino masses.
In the present work we substantiate further this statement, by making an
analytic study of the rates for one of the processes discussed in
ref.~\cite{BH}, $\mu\to e\gamma$, in the full parameter space of the
unified theory, both in~SU(5) \cite{GG} and in~SO(10) \cite{GFM} and by
subsequently comparing it with the other processes and quantities of
interest.
In addition to generating lepton flavour violating interactions, the large
top quark Yukawa coupling also leads to important contributions to the
electric dipole moment of the electron and neutron in~SO(10) theories.
In this paper we give the electric dipole moment over the full parameter
space of the~SO(10) unified theory.

The paper is organized as follows.
In section~2 we discuss the physical
mechanism which allows the top quark Yukawa coupling to generate large
amplitudes for processes which violate individual lepton numbers.
In section~3 we summarize the present information on
$\lambda_{t\rm G}$ and we describe an upper bound on
$\lambda_{t\rm G}$ arising from the infrared fixed point
behaviour of the top Yukawa coupling
above the unification scale up to $M_{\rm Pl}$.
In section~4 we study the scaling of the supersymmetry breaking
parameters in~SU(5) with emphasis on the flavour violating effects
due to $\lambda_{t\rm G}$.
In section~5 we give, in~SU(5), the pieces of the low energy
Lagrangian relevant to the calculation of the~LFV processes
in the physical lepton and slepton basis.
In section~6 we calculate the rate for $\mu\to e\gamma$ in the
full space of parameters.
In sections~7--9 we extend the analysis of sections~4--6 to the~SO(10) case.
In section~10 we discuss the $\mu\to e$ conversion in atoms
and we establish the relative merit of the study of this
process with respect to $\mu\to e\gamma$ in the search for a signal
of lepton flavour violation.
The same is done in section~11 for the $\tau\to\mu\gamma$ decay.
Finally, in section~12, we study the relation between $\mu\to e\gamma$
and the electric dipole moment of the electron~\cite{DH}.
Our conclusions are drawn in section~13.
Appendices~A and~B contain the analytic solutions of all the
relevant Renormalization Group Equations
from $M_{\rm Pl}$ to $M_{\rm G}$ (appendix~A) and
from $M_{\rm G}$  to $M_Z$ (appendix~B), both in~SU(5) and in~SO(10).

\section{The origin of lepton flavour violation}
In this paper we study grand unified theories which incorporate weak-scale
supersymmetry~\cite{DG} and have the origin of supersymmetry breaking near
at the Planck scale~\cite{BFS}.
These theories lead to the successful weak mixing angle prediction, and, as a
promising direction for unifying both the forces and the
fundamental fermions, are currently receiving much attention.
In all such theories, we find that the large top quark
Yukawa coupling leads to a rate for $\mu \rightarrow e \gamma$
which can be reliably computed in terms of weak-scale parameters~\cite{BH}.
Over much of the interesting parameter space,
the rate is within two orders of magnitude of the
present experimental limit.
At first sight, it is surprising that the top quark Yukawa coupling should lead
to any violation of $L_e$ or $L_\mu$. What is the physical origin of this
effect, and why is it not suppressed by inverse powers of $M_{\rm G}$? The
answer
lies in new flavour mixing matrices, which are analogous to the
Cabibbo-Kobayashi-Maskawa matrix.

In the standard model the quark mass eigenstate basis is reached by making
independent rotations on the left-handed up and down type quarks, $u_L$
and $d_L$.
However, these states are unified into a doublet of the weak SU(2) gauge
group: $Q = (u_L, d_L)$.
A relative rotation between $u_L$ and $d_L$ therefore
leads to flavour mixing at the charged $W$ gauge vertex.
This is the well-known
Cabibbo-Kobayashi-Maskawa mixing. With massless neutrinos,
the standard model has no analogous flavour mixing amongst the leptons:
the charged lepton mass eigenstate basis can be reached by a rotation of
the entire lepton doublet $L=(\nu_L, e_L)$.

How are these considerations of flavour mixing altered in supersymmetric
unified theories? There are two new crucial ingredients.
The first is provided by weak-scale supersymmetry, which implies that
the quarks and leptons have scalar partners.
The mass eigenstate basis for these squarks and sleptons requires
additional flavour rotations. As an example, consider softly broken
supersymmetric QED with three generations of charged leptons.
There are three arbitrary mass matrices, one for the charged leptons,
$e$, and one each for the left-handed and right-handed sleptons,
$\tilde{e}_L$ and $\tilde{e}_R$.
To reach the mass basis therefore requires relative rotations between
$e_L$ and $\tilde{e}_L$ as well as between $e_R$ and $\tilde{e}_R$,
resulting in two flavour mixing matrices at the photino gauge vertex.

In supersymmetric extensions of the standard model, these additional
flavour-changing effects are known to be problematic. With a mixing angle
comparable to the Cabibbo angle, a branching ratio for  $\mu \rightarrow
e \gamma$ of order $10^{-4}$ results. In the majority of
supersymmetric models which have been constructed, such flavour-changing
effects have been suppressed by assuming that the origin of supersymmetry
breaking is flavour blind. In this case the slepton
mass matrix is proportional to the unit matrix.
The lepton mass matrix can then be diagonalized by
identical rotations on $e_L$ and $\tilde{e}_L$ as well as on
$e_R$ and $\tilde{e}_R$, without introducing
flavour violating mixing matrices at the gaugino vertices.
{\it Slepton degeneracy renders lepton
flavour mixing matrices non-physical.}

The unification of quarks and leptons into larger multiplets provides the
second crucial new feature in the origin of flavour mixing~\cite{HKR}.
The weak unification of $u_L$ and $d_L$ into $Q$ is extended in SU(5)
to the unification of $Q$ with
$u^c_L$ and $e^c_L$ into a 10 dimensional multiplet $T(Q,u^c_L,e^c_L)$.
Since higher unification leads to fewer multiplets, there are fewer
rotations which can be made without generating flavour mixing matrices.

In any supersymmetric unified model there must be at least two coupling
matrices, $\mb{\lambda}_1$ and $\mb{\lambda}_2$, which describe quark
masses.
If there is only one such matrix, it can always be diagonalized without
introducing quark mixing. One of these coupling matrices, which we take to
be $\mb{\lambda}_1$, must contain the large coupling, $\lambda_t$, which is
responsible for the top quark mass. We choose to work in a basis in which
$\mb{\lambda}_1$ is diagonal. The particles which interact via $\lambda_t$
are those which lie in the same unified multiplet with the top. In all
unified models this includes a right-handed charged lepton, which we call
$e_{L_3}^c$. This cannot be identified as the mass eigenstate $\tau_L^c$,
because significant contributions to the charged lepton masses must come
from the matrix $\mb{\lambda}_2$, which is not diagonal.

The assumption that the supersymmetry breaking mechanism is flavour blind
leads to mass matrices for both $\tilde{e}_L$ and $\tilde{e}_R$ which are
proportional to the unit matrix at the Planck scale, $M_{\rm Pl}$.
As we have seen, without unified interactions,
lepton superfield rotations can diagonalize the
lepton mass matrix without introducing flavour mixing matrices.
However, the unification prevents such rotations: the leptons are in
the same multiplets as quarks, and the basis has already been chosen to
diagonalize $\mb{\lambda}_1$.
As the theory is renormalization group scaled to lower energies, the
$\lambda_t$ interaction induces radiative corrections which suppress
the mass of $\tilde{e}_{R_3}$ beneath that of
$\tilde{e}_{R_2}$ and $\tilde{e}_{R_1}$.
Beneath $M_{\rm G}$ the superheavy particles of the theory can be
decoupled, leaving only the interactions
of the minimal supersymmetric standard model.
Now that the unified symmetry which relates quarks to leptons is broken,
a lepton mass
basis can be chosen by rotating lepton fields relative to quark fields.
However, at these lower energies the sleptons are no longer degenerate,
so that these rotations do induce lepton flavour mixing angles.
{\it Radiative corrections induced by $\lambda_t$ lead to slepton
non-degeneracies,
which render the lepton mixing angles physical}~\cite{BH}.

This discussion provides the essence of the physics mechanism for lepton
flavour violation in superunified models. Since the flavour mixing
matrices have complex entries, they also lead to~CP violation.
It shows the effect to be generic to the idea of quark-lepton unification,
requiring only that the superpartners have masses around the Fermi scale,
and that supersymmetry breaking be present at the Planck scale.
The imprint of the unified interactions is made on the soft supersymmetry
breaking coefficients, including the scalar trilinears, which are taken to
be flavour blind at the Planck scale. Eventually this imprint will be seen
directly by studying the superpartner spectrum, but it can also be probed
now by searching for $L_e$, $L_\mu$, $L_\tau$ and CP violating effects.

\section{The top Yukawa coupling at the GUT scale}
The top Yukawa coupling at the unification scale, $\lambda_{t\rm G}$,
plays a crucial role in the determination of the~LFV effects
discussed in this paper.
In this section we therefore summarize the present information on
$\lambda_{t\rm G}$ which comes from two different sources: the direct
measurement of the top mass and, indirectly, the bottom/tau mass ratio.

The top Yukawa coupling $\lambda_{t\rm G}$ can of course be easily scaled
down to determine its value at the weak scale $\lambda_t$
(see eq.~(\ref{eq:lambda_t}) of app.~\ref{RGEMSSM}).
In turn, $\lambda_t$ determines the top quark pole mass via~\cite{TopPole}
\begin{equation}\label{eq:m_t}
M_t=\lambda_t\, v\sin\beta\left(1+{4\over3}{\alpha_3(M_t)\over\pi}+
11.4{\alpha_3^2\over\pi^2}\right),\qquad v=174\,{\rm GeV},
\end{equation}
where, as usual, $\tan\beta=v_{\rm u}/v_{\rm d}$ is the ratio
of the two light Higgs vacuum expectation values.
Figure~\ref{fig:ltmaxa3} shows
$\lambda_{t\rm G}$ as function of the strong coupling constant
$\alpha_3(M_Z)$, for $M_t=174\pm16\,{\rm GeV}$ and for moderate
($\tan\beta=2$) or relatively high values ($\tan\beta=10$)
of the ratio $v_{\rm u}/v_{\rm d}$.
The rapid saturation for large $\tan\beta$ implies that the lowest
curve in fig.~\ref{fig:ltmaxa3} is actually a lower bound on
$\lambda_{t\rm G}$ for $m_t>158\,{\rm GeV}$.

As is well known, the Yukawa superpotential of
minimal~SU(5) allows a prediction for the ratio $m_b/m_\tau$ as a function
of $\lambda_{t\rm G}$ and $\alpha_3(M_Z)$~\cite{btau1,btau2}.
This prediction is given for the running $b$-mass $m_b(m_b)$ in
fig.~\ref{fig:b-tau}, and is
compared with the preferred value as determined from $\Upsilon$-physics.
The dependence on $\tan\beta$ drops out in the ratio $m_b/m_\tau$,
except for $\tan\beta\approx m_t/m_b$.
For moderate values of $\tan\beta$ there is a clear consistency between
fig.~\ref{fig:ltmaxa3} and fig.~\ref{fig:b-tau} with a strong indication
for a high value of $\lambda_{t\rm G}$.
The consistency is weaker for larger $\tan\beta$, unless the top mass
is close to $200\,{\rm GeV}$, in the upper range of preliminary values
indicated by the~CDF experiment~\cite{expTop}.
In this case, of course, a rather high value of $\lambda_{t\rm G}$
is also indicated, resulting in a large flavour violation
of the soft supersymmetry breaking parameters at the unification scale.
For very large $\tan \beta$, close to $m_t/m_b$, consistency with $m_b/m_\tau$
is possible even for the smallest values of $\lambda_{t\rm G}$ allowed by the
top mass.
We have not studied this case in this paper. However, because the
$\mu \to e \gamma$ amplitude always contains a term proportional to
$\tan\beta$, the rate is always significant for such large values of
$\tan\beta$.

For later purposes, it will be useful to know the behaviour of
$\lambda_t$ at energies above $M_{\rm G}$, as determined from the
Renormalization Group Equations (RGE).
Assuming that the unified gauge coupling $g_5$ and $\lambda_t$
itself are the only relevant couplings, the~RGEs are solved
in appendix~A at the one loop level.
The solution for $\lambda_t$ displays an infrared fixed point. If
$\lambda_t(M_{\rm Pl})$ is large, but still perturbative,
it will be drawn to the infrared fixed point at $M_{\rm G}$.
The value of the coupling at the fixed point at
$M_{\rm G}$ is larger for~SU(5) than for~SO(10), and depends on
the one loop coefficient, $b_{\rm G}$, of the gauge $\beta$-function,
as shown in figure~\ref{fig:ltmaxbG}.
The quantity  $\lambda^{\rm max}_{t\rm G}$ plotted in this
figure is the value of  $\lambda_{t\rm G}$ for which the one loop evolved
value of  $\lambda_t(M_{\rm Pl})$ becomes infinite.
For all the numerical work of this paper we take
$\lambda_{t\rm G} < \lambda^{\rm max}_{t\rm G}$,
so that perturbation theory can be trusted.
For larger values of  $\lambda_{t\rm G}$ the theory becomes
non-perturbative at scales beneath $M_{\rm Pl}$.
Although we are unable to make
computations for this case, the non-perturbative coupling is expected to
generate large non-degeneracies amongst the scalars, leading to large
rates for $\mu \to e \gamma$.

\section{Scaling of supersymmetry breaking parameters in SU(5)}
The messengers of flavour violation in the lepton sector are
the soft supersymmetry breaking
terms, which are therefore crucial to determine.
Without having to specify the actual mechanism of supersymmetry
breaking, nor the sector in which it takes place, we assume that it is
transmitted to standard matter by supergravity couplings~\cite{BFS} and
that it results, at the Planck scale, in universal soft breaking terms.

Standard matter occurs in the usual triplication of
$10~(T)\oplus \bar{5}~(\bar{F})$ representations of~SU(5), which are
coupled to a $5~(H)$ and a $\bar{5}~(\bar{H})$ representation of
Higgs supermultiplets in the Yukawa superpotential
\begin{equation}\label{eq:WYSU(5)}
W=T_i    \lambda ^{\rm u}_{ij} T_j H +
  T_i    \lambda ^{\rm d}_{ij}\bar{F}_j\bar{H}\equiv
  T^T\mb{\lambda}^{\rm u} T H +
  T^T\mb{\lambda}^{\rm d}\bar{F}\bar{H}
\end{equation}
which we assume to be valid from $M_{\rm G}$ to the Planck scale.
The full superpotential will contain other supermultiplets $\Sigma$,
needed to break~SU(5) but not directly coupled to matter.
Assuming no large Yukawa couplings of the $\Sigma$ fields to the
$H,\bar{H}$ multiplets, the $\Sigma$ fields affect the determination of the
soft supersymmetry breaking terms at the~GUT scale only via their
contribution to the gauge $\beta$-function
from $M_{\rm G}$ to $M_{\rm Pl}$.
Unless otherwise specified, we shall take the~SU(5) $\beta$-function
coefficient of the minimal Dimopoulos-Georgi model \cite{DG}.
Different $\beta$-function coefficients mostly
affect the rates for the~LFV processes only through the restrictions
that they induce on the range of the low energy parameters
(see appendix~\ref{RGE5}).

The relevant part of the soft supersymmetry breaking Lagrangian,
before~SU(5) breaking, has the form
\begin{equation}\label{eq:Lsoft5}
-\Lag_{\rm soft}=V_{\rm soft}=
      \tilde{T}^\dagger \mb{m}_T^2\tilde{T}+
\tilde{\!\bar{F}}^\dagger \mb{m}_{\bar{F}}^2 \tilde{\!\bar{F}}+
m_H^2 |H|^2 + m_{\bar{H}}^2 |\bar{H}|^2+
\tilde{T}^T \mb{A}^{\rm u}\mb{\lambda}^{\rm u}\tilde{T} H+
\tilde{T}^T \mb{A}^{\rm d}\mb{\lambda}^{\rm d}\tilde{\!\bar{F}} \bar{H}
\end{equation}
with, at the Planck scale,
\begin{equation}\label{eq:m0A0}
\begin{array}{c}
\mb{m}_T^2 = \mb{m}_{\bar{F}}^2 = m_0^2 \One,\\[1mm]
m_H^2=m_{\bar H}^2 = m_0^2,\\[1mm]
\mb{A}^{\rm u} = \mb{A}^{\rm d} = A_0 \One.
\end{array}\end{equation}
The renormalization of the parameters in~(\ref{eq:Lsoft5})
down to the~GUT scale is most easily done by working in the basis where
the Yukawa matrix $\mb{\lambda}^{\rm u}$ has diagonal form
(hereafter called the u-basis).
By keeping in the RGE only the
one loop effects due to the~SU(5) gauge coupling and to the third
entry of $\mb{\lambda}^{\rm u}$, $\lambda^{\rm u}_{33}=\lambda_t$,
it is simple to rescale down to $M_{\rm G}$ the soft breaking
parameters (see appendix~\ref{RGE5}).
Flavour universality is of course no longer maintained.
In fact, the mass term for the ten-plet and the $A$-terms acquire
the form
\begin{eqnsystem}{sys:softGUT5}
\mb{m}_{T\rm G}^2 &=&
\diag(m_{T\rm G}^2,m_{T\rm G}^2,m_{T\rm G}^2-I_{\rm G})
\equiv m_{T\rm G}^2\One - \I_{\rm G},\\
\mb{A}^{\rm d}_{\rm G} &=&\textstyle
\diag(A_{d\rm G},A_{d\rm G},A_{d\rm G}- \frac{1}{3}I'_{\rm G})
\equiv A_{d\rm G}\One-\frac{1}{3}\I'_{\rm G},\\
\mb{A}^{\rm u}_{\rm G} &=&\textstyle
\diag(A_{u\rm G}-\frac{1}{3}I'_{\rm G},A_{u\rm G}-\frac{1}{3}I'_{\rm G}
,A_{u\rm G}-I'_{\rm G}).
\end{eqnsystem}
At the same time, the mass matrix of the
five-plets maintains the universal form
\begin{equation}
\mb{m}_{\bar{F}\rm G}^2=m_{\bar{F}\rm G}^2\One
\end{equation}
and the Yukawa coupling matrix $\mb{\lambda}^{\rm u}$ remains diagonal,
\begin{equation}
\mb{\lambda}^{\rm u}_{\rm G}=
\diag(\lambda_{u\rm G},\lambda_{c\rm G},\lambda_{t\rm G}),
\end{equation}
whereas $\mb{\lambda}^{\rm d}$ gets renormalized
to      $\mb{\lambda}^{\rm d}_{\rm G}$.
The explicit expressions for
$m_{T\rm G}^2$, $m_{\bar{F}\rm G}^2$,
$A_{d\rm G}$, $A_{u\rm G}$, $I_{\rm G}$ and $I'_{\rm G}$, as well as
the renormalization of the Higgs mass parameters
are given in appendix~\ref{RGE5}.
The flavour breaking parameters $I_{\rm G}$ and $I'_{\rm G}$
have a crucial dependence
on the top Yukawa coupling at $M_{\rm G}$, $\lambda_{t\rm G}$
(see section~5).

In this paper we take a universal boundary condition for the soft
supersymmetry breaking parameters at the Planck scale. How do our results
depend on this assumption? A well motivated relaxation of this assumption
is to allow soft scalar masses to be the most general allowed by the gauge
symmetry and by a symmetry which interchanges one generation with another.
This would satisfy flavour changing phenomenology without forcing
identical Higgs and matter scalar masses, and would also allow the scalars
in $T$ to have masses different from those in $\bar{F}$. Although extra
parameters must be introduced, this generalization will not affect our
results in a crucial way. More important would be the addition of small
flavour changing scalar masses at the Planck scale, since they would lead
directly to the processes which we discuss in this paper. These
contributions would simply add to those which we calculate here. While
cancellations cannot be excluded, we believe they would have to be
accidental. For example, the contributions from the Planck scale boundary
condition would arise from string physics and would be independent of the
value of $M_{\rm G}$. On the other hand, the contributions calculated in
this paper do depend on $M_{\rm G}$.
\begin{figure}
\caption{The top Yukawa coupling at $M_{\rm G}$ for
$\tan\beta=2$ (full lines) and $\tan\beta=10$ (dashed lines)
for $M_t=158,174,192\,{\rm GeV}$ (in increasing order),
as function of $\alpha_3(M_Z)$.\label{fig:ltmaxa3}}
\caption{The running $b$-quark mass in the
$\alpha_3(M_Z), \lambda_{t\rm G}$ plane from $b/\tau$ unification.
The darker area corresponds to $m_b(m_b)=4.25\pm0.10\,{\rm GeV}$,
as obtained from $\Upsilon$-physics.\label{fig:b-tau}}
\caption{Fixed point upper bounds on
$\lambda_{t\rm G}$ in~SU(5) and~SO(10), as defined
in the text, as functions of the one loop coefficient $b_{\rm G}$
of the gauge $\beta$-function.\label{fig:ltmaxbG}}
\end{figure}

\section{The low energy Lagrangian in SU(5)}
After SU(5) breaking, the scaling down to low energy of the various
parameters results in the low energy Lagrangian, whose relevant pieces
are summarized for ease of the reader.
They are, to first order in the Yukawa couplings:
\begin{itemize}
\item[i.] The slepton mass matrix
\begin{equation}\label{eq:lowEnLag}
-\Lag_m^{\rm sl}=
\tilde{L}^\dagger \mb{m}_L^2\tilde{L} +
\tilde{e}_R^\dagger \mb{m}_e^2 \tilde{e}_R +
\tilde{e}_R^T (\mb{A}^{\rm e} +  \One\mu\tan\beta)\mb{\lambda}^{\rm e}
\tilde{e}_L v_{\rm d}+{\rm h.c.}
\end{equation}
where $\tilde{L}$, $\tilde{e}_R$ are 3-vectors containing
the~SU(2) doublet and singlet sleptons,
\begin{equation}
\mb{m}_L^2=m_L^2\One,\qquad
\mb{m}_e^2=m_e^2\One-\I_{\rm G},\qquad
\mb{A}^{\rm e}=A_e\One-{\textstyle\frac{1}{3}}\I'_{\rm G},
\end{equation}
$m_L^2$, $m_e^2$ and $A_e$ are given in appendix~\ref{RGEMSSM},
and a term proportional to the $\mu$ parameter has been
explicitly introduced;
\item[ii.] the Higgs mass terms
\begin{equation}\label{eq:mumdmud}
-\Lag_m=
(m_{\rm u}^2+\mu^2) |h_{\rm u}|^2 +
(m_{\rm d}^2+\mu^2) |h_{\rm d}|^2 -
m_{\rm ud}^2 (h_{\rm u} h_{\rm d} + {\rm h.c.})
\end{equation}
with $m_{\rm u}^2$, $m_{\rm d}^2$
given in eq.~(\ref{sys:mumd}) of appendix~\ref{RGEMSSM}.

\item[iii.] the quarks and lepton mass terms
\begin{equation}
\Lag_Y=
     Q^T\mb{\lambda}^{\rm u}_Z u^c_L\cdot v_{\rm u} +
     Q^T\mb{\lambda}^{\rm d}_Z d^c_L\cdot v_{\rm d} +
e^{cT}_L\mb{\lambda}^{\rm e}_Z L    \cdot v_{\rm d}
\end{equation}
where, in the u-basis, $\mb{\lambda}^{\rm u}_Z$ has kept
its diagonal form and the matrices
$\mb{\lambda}^{\rm d}$ and $\mb{\lambda}^{\rm e}$,
equal at $M_{\rm G}$, have been shifted by the different renormalization
effects due to $\lambda_t$ and the gauge couplings.
\end{itemize}
The LFV parameter $I_{\rm G}$ is directly related to the splitting
between the $\tilde{\tau}_R$ and the $\tilde{e}_R$ ($\tilde{\mu}_R$).
The $\tilde{\tau}_R$-mass is shown in figures~\ref{fig:mstauR}
for fixed values of $\lambda_{t\rm G}$ and of the
$\tilde{e}_R$-mass, as function of $A_e$ and of the wino mass
$M_2$ in its full range, as determined from $m_{\tilde{e}_R}$
itself.
We take the value of $\lambda_{t\rm G}$ such that
$\lambda_{t\rm G}^2=0.8(\lambda_{t\rm G}^{\rm max})^2$.
The lightness of $\tilde{\tau}_R$ is of course a main consequence of
the present picture as far as the superpartner spectrum
is concerned.
Another interesting consequence is the strong upper bound on the
gaugino mass for any given $m_{\tilde{e}_R}$, which results in particular
in the lightest supersymmetric particle being always the lightest
neutralino.
The $\tilde{\tau}_R$-mass has a negligible dependence on
$\tan\beta$ for $m_{\tilde{e}_R}^2\gg M_Z^2$.
For values of $m_{\tilde{e}_R}$ higher than~300~GeV,
$m_{\tilde{\tau}_R}$ and $M_2$ rescale in the same way as
$m_{\tilde{e}_R}$ itself does.

By diagonalizing $\mb{\lambda}^{\rm d}_Z$ and
$\mb{\lambda}^{\rm e}_Z$, we have
\begin{eqnsystem}{sys:UVde}
\mb{\lambda}^{\rm d}_Zv_{\rm d}&=&\mb{V}^* \mb{M}^{\rm d}\mb{U}^\dagger\\
\mb{\lambda}^{\rm e}_Zv_{\rm d}&=&\mb{V}^{\rm e*}\mb{M}^{\rm e}
\mb{U}^{\rm e\dagger}\label{eq:Ue}
\end{eqnsystem}
where $\mb{M}^{\rm d}$, $\mb{M}^{\rm e}$ are the diagonal mass
matrices for down quarks and charged leptons,
$\mb{U}=\mb{U}^{\rm e}$,
$\mb{V}$ is the usual Cabibbo-Kobayashi-Maskawa matrix and,
as an effect of the top Yukawa coupling, the matrix elements of
$\mb{V}^{\rm e}$ are related to those of $\mb{V}$ by~\cite{Vevol}
\begin{equation}\label{eq:Vscaling}
V^{\rm e}_{ij}=y V_{ij}\quad\hbox{for $i\neq j$ and ($i$ or $j)=3$},
\qquad
V^{\rm e}_{ij}= V_{ij}\quad{\rm otherwise}
\end{equation}
and $y$ is defined in eq.~(\ref{eq:y}) of app.~B.
We ignore for the time being the fact that one does not obtain
in this way the correct relation between the masses of the light
leptons and down quarks, which are also related to each other
by an appropriate renormalization group rescaling.

It is convenient to work in a mass eigenstate basis
for the charge leptons, which is simply obtained by the redefinitions
(with primed indices suppressed after eq.~(\ref{eq:ridef}))
\begin{equation}\label{eq:ridef}
\mb{V}^{\rm e\dagger} e^c_L = e^{\prime c}_L,\qquad
\mb{U}^{\rm e\dagger} L = L'.
\end{equation}
In the gaugino couplings, the rotation on the charged lepton doublets
can be compensated by the same rotation of the full supermultiplets,
since the SU(2) doublet slepton mass matrix has kept its diagonal form
through renormalization,
whereas this is not the case for the singlets $\tilde{e}_R$.
As a consequence, the matrix $\mb{V}^{\rm e}$
appears in the neutralino couplings
\begin{equation}\Lag_g=\sqrt{2}g'\sum_{n=1}^4\Big[-{1\over2}
\overline{e_L}\,\tilde{e}_L N_n
(H_{n\tilde{B}}+\cot\theta_{\rm W}H_{n\tilde{W}_3})+
\overline{e^c_L}\,\mb{V}^{\rm e\dagger}\tilde{e}_R N_n H_{n\tilde{B}}+
{\rm h.c.}\Big]
\end{equation}
where $N_n$ are the four neutralino mass eigenstates, of mass $M_n$,
related to the bino and the neutral wino by
\begin{equation}
\begin{array}{c}
\tilde{B}  = \sum_{n=1}^4 N_n H_{n\tilde{B}}, \\[2mm]
\tilde{W}_3= \sum_{n=1}^4 N_n H_{n\tilde{W}_3}.
\end{array}
\end{equation}
Notice that, in the slepton basis in which we are working,
also the third term in the right side of~(\ref{eq:lowEnLag}) has
non diagonal form, being
\begin{equation}
-\Lag_m^{\rm n.d.}=
(A_e + \mu\tan\beta)
\tilde{e}_R^T    \mb{V}^{\rm e*}\mb{M}^{\rm e}\tilde{e}_L-
\tilde{e}_R^T {\textstyle\frac{1}{3}}
\I'_{\rm G}\mb{V}^{\rm e*}\mb{M}^{\rm e}\tilde{e}_L+{\rm h.c.}
\end{equation}

\begin{figure}
\caption[mstauR]{Isoplots of $m_{\tilde{\tau}_R}$
in the $M_2,A_e/m_{\tilde{e}_R}$ plane for
(a) $\lambda_{t\rm G}=1.4$, $m_{\tilde{e}_R}=100\,{\rm GeV}$,
 $\tan\beta=2$ in SU(5),
(b) $\lambda_{t\rm G}=1.4$, $m_{\tilde{e}_R}=100\,{\rm GeV}$,
 $\tan\beta=10$ in SU(5),
(c) $\lambda_{t\rm G}=1.4$, $m_{\tilde{e}_R}=300\,{\rm GeV}$,
 $\tan\beta=2,10$ in SU(5), or, for $M_2<270\,{\rm GeV}$,
$\lambda_{t\rm G}=1.25$, $m_{\tilde{e}_R}=300\,{\rm GeV}$,
 $\tan\beta=2,10$ in SO(10),
(d) $\lambda_{t\rm G}=0.86$, $m_{\tilde{e}_R}=300\,{\rm GeV}$,
 $\tan\beta=2,10$ in SO(10).
In fig.s~\ref{fig:mstauR}a,b the isolines are separated by 10~GeV,
in fig.s~\ref{fig:mstauR}c,d by 30~GeV.\label{fig:mstauR}}
\caption{Lepton flavour violating couplings in SU(5).\label{fig:Ver5}}
\caption[figuraSei]{Diagrams giving rise to the decay
$\mu\to e\gamma$ in SU(5).
In figures~\ref{fig:GraMu5}b,c, an external photon line is left
understood, which can be attached to either of the scalar lines.
\label{fig:GraMu5}}
\caption{Isoplots of ${\rm B.R.}(\mu\to e\gamma)$ in SU(5) in the
$M_2,A_e/m_{\tilde{e}_R}$ plane for $\lambda_{t\rm G}=1.4$,
$m_{\tilde{e}_R}=100\,{\rm GeV}$ and
(a) $\tan\beta=2 ,~\mu<0$,
(b) $\tan\beta=2 ,~\mu>0$,
(c) $\tan\beta=10,~\mu<0$,
(d) $\tan\beta=10,~\mu>0$.
The dashed (dotted) lines delimit regions where
$m_{\tilde{\tau}_R}^2<0$ ($\mu^2<0$).
The shaded area also extends to
$m_{\tilde{\tau}_R}<45\,{\rm GeV}$.
The darker area shows a region where the rate is small,
and passes through zero, due to a cancellation of terms.
The dot-dashed line corresponds to the present experimental limit.
For the CKM matrix elements we take $|V_{cb}|=0.04$ and $|V_{td}|=0.01$.
\label{fig:mU100}}
\caption[Same]{Same as in fig.~\ref{fig:mU100} for
$m_{\tilde{e}_R}=300\,{\rm GeV}$.\label{fig:mU300}}
\end{figure}

\section{$\mu\to e\gamma$ in supersymmetric SU(5)}
The LFV couplings are summarized in fig.~\ref{fig:Ver5}.
Correspondingly, if we neglect the electron mass and we work to first
order in $m_e/m_\mu$ or $m_\mu/m_\tau$,
the diagrams giving rise to the decay
$\mu\to e\gamma$ are shown in fig.~\ref{fig:GraMu5}.
Taking into account that the selectron, $\tilde{e}_R$,
and smuon, $\tilde{\mu}_R$, singlets are
degenerate at a common squared mass $m_{\tilde{e}_R}^2$,
whereas they are split from the stau singlet $\tilde{\tau}_R$,
of squared mass $m_{\tilde{\tau}_R}^2=m_{\tilde{e}_R}^2-I_{\rm G}$,
and using the unitarity of the matrix $\mb{V}^{\rm e}$,
one obtains the following contributions to the $\mu\to e\gamma$
decay amplitude
\begin{equation}\label{eq:AmuSU5}
{\cal A}_\mu(\mu\to e\gamma)=
-ie\cdot\bar{u}_e i \sigma_{\mu\nu} q^\nu
\frac{1-\gamma_5}{2} u_\mu\, F_2
\end{equation}
\begin{eqnsystem}{sys:F2abc}
F_2^{\rm (a)} &=& \frac{\alpha}{4\pi\cos^2\theta_{\rm W}}
m_\mu V_{\tau\mu}^{\rm e*}V_{\tau e}^{\rm e}
[G_1(m_{\tilde{\tau}_R}^2)-G_1(m_{\tilde{e}_R}^2)]\\
\noalign{\hbox{~b.~~from the diagram of figure~\ref{fig:GraMu5}b:}}
\nonumber\\
F_2^{\rm (b)} &=& \frac{\alpha}{4\pi\cos^2\theta_{\rm W}}
m_\mu V_{\tau\mu}^{\rm e*}V_{\tau e}^{\rm e}(A_e+\mu\tan\beta)
[G_2(m_{\tilde{e}_L}^2,m_{\tilde{\tau}_R}^2)-
 G_2(m_{\tilde{e}_L}^2,m_{\tilde{e}_R}^2)]\\
\noalign{\hbox{~c.~~from the diagram of figure~\ref{fig:GraMu5}c:}}
\nonumber\\
F_2^{\rm (c)} &=& \frac{\alpha}{4\pi\cos^2\theta_{\rm W}}
m_\mu V_{\tau\mu}^{\rm e*}V_{\tau e}^{\rm e}
(-{\textstyle\frac{1}{3}}I'_{\rm G})
G_2(m_{\tilde{e}_L}^2,m_{\tilde{\tau}_R}^2)
\end{eqnsystem}
where
\begin{eqnarray*}
G_1(m^2) &=& \sum_{n=1}^4 \frac{H_{n\tilde{B}}^2}{M_n^2}
g_1(\frac{m^2}{M_n^2}),\qquad~~
g_1(r)={-1\over6(r-1)^4}[2+3r-6r^2+r^3+6r\ln r]\\
\noalign{\hbox{and}}\\
G_2(m^2) &=& \sum_{n=1}^4 \frac{H_{n\tilde{B}}}{M_n}
(H_{n\tilde{B}}+\cot\theta_{\rm W}H_{n\tilde{W}_3})
\cdot g_2(\frac{m^2}{M_n^2}),\\
G_2(m_1^2,m_2^2) &=& \frac{G_2(m_1^2)-G_2(m_2^2)}{m_1^2-m_2^2},\qquad
g_2(r)={1\over2(r-1)^3}[r^2-1-2r\ln r].
\end{eqnarray*}
Correspondingly, the decay rate is given by\footnote{
In the limit of small $I_{\rm G}$, this rate agrees with the analytic
expression given in ref.~\cite{BH}.
However, in view of the values
actually taken by $I_{\rm G}$, the expansion generally gives a poor
approximation to the correct rate.
Previous calculations of the $\mu\to
e\gamma$ rate for special values of the gaugino and slepton masses were
made in references~\cite{Top}.}
\begin{equation}\label{eq:GammaSU5}
\Gamma(\mu\to e\gamma)=\frac{\alpha}{4} m_\mu^3 |F_2|^2,\qquad
F_2 = F_2^{\rm (a)} + F_2^{\rm (b)} + F_2^{\rm (c)}.
\end{equation}
Equations~(\ref{sys:F2abc},\ref{eq:GammaSU5}), together with the
expressions of the parameters in the low energy Lagrangian
as defined in the previous section
and explicitly given in appendix~\ref{RGEMSSM},
allow the numerical calculation of the branching ratio
${\rm B.R.}(\mu\to e\gamma)$, shown in fig.~\ref{fig:mU100}
and fig.~\ref{fig:mU300}
for $m_{\tilde{e}_R}$ equal to $100\,{\rm GeV}$ and $300\,{\rm GeV}$
respectively. For values of $m_{\tilde{e}_R}$ greater than $300\,{\rm GeV}$
and fixed $A_e/m_{\tilde{e}_R}$, $M_2/m_{\tilde{e}_R}$ the branching
ratio scale as $m_{\tilde{e}_R}^{-4}$.
{}From~(\ref{eq:Vscaling}), for the CKM matrix elements we take
$|V_{cb}|=0.04$ and $|V_{td}|=0.01$.

The set of independent parameters, on which the branching ratio depends,
are $\{A_0,m_0^2,M_{5\rm Pl}\}$ which determine the soft operators,
the top quark Yukawa coupling,
the coefficient of the one-loop gauge beta function of the
    unified theory, $b_{\rm G}$,
the ratio of weak vacuum expectation values, $\tan\beta$, and
the Higgs mixing parameter $\mu$.
We choose to exchange $\{A_0,m_0^2,M_{5\rm Pl}\}$ for the physically more
interesting set $\{A_e,m_{\tilde{e}_R}^2,M_2\}$,
where $A_e$ is the light generation lepton $A$-parameter,
$m_{\tilde{e}_R}$ is the mass of the right-handed selectron and
$M_2$ is the weak scale gaugino mass parameter for~SU(2).
In appendix~A the full dependence of quantities on $b_{\rm G}$ is given,
and is found to be mild,
hence we have chosen the minimal value $b_{\rm G} = -3$.
The $\mu$ parameter which enters in $F_2^{\rm (b)}$,
eq.~(\ref{sys:F2abc}b), and also in the neutralino
mass matrix, is expressed, up to its sign, in terms of
the other parameters by means of the electroweak symmetry
breaking relation
\begin{equation}
\mu^2=-{M_Z^2\over2} -
\frac{m_{\rm d}^2-m_{\rm u}^2\tan^2\beta}{1-\tan^2\beta}
\end{equation}
with $m_{\rm u}^2$, $m_{\rm d}^2$ given in appendix~\ref{RGEMSSM}.

\section{Scaling of supersymmetry breaking parameters in SO(10)}
In the case of SO(10) gauge symmetry, in fact as one
of its most attractive features,
the quarks and leptons of a single generation are the components
of a single 16-dimensional spinorial representation $\Psi$.
This is a crucial feature for the problem at hand;
it causes all the scalars of the third
generation, and not only those in the~10 of~SU(5), to be lighter
than the corresponding scalars in the first and the second generation.
In turn, and at variance with~SU(5), LFV interactions arise
also involving the left handed sleptons.
With this in mind, the considerations of the previous sections can be
straightforwardly extended to the~SO(10) case, after specifying
the Yukawa superpotential and the gauge $\beta$-function,
at one loop, at the Planck scale.
For simplicity we assume that~SO(10) is broken at once to the
low energy standard group at $M_{\rm G}$.

In SO(10) gauge theories a single Yukawa interaction of the three
spinorial matter multiplets $\Psi_i$ to a vector 10-dimensional
Higgs representation $\Phi$, $\Psi^T\mb{\lambda}\Psi\Phi$,
does not describe any intergenerational mixing, since $\Psi$ can
be rotated to make $\mb{\lambda}$ diagonal.
To describe the mixing, we introduce two 10-plets,
$\Phi_{\rm u}$ and $\Phi_{\rm d}$, in the superpotential \cite{DH}
\begin{equation}
W_{\rm SO(10)}=\Psi^T\mb{\lambda}^{\rm u}\Psi\Phi_{\rm u}+
\Psi^T\mb{\lambda}^{\rm d}\Psi\Phi_{\rm d}
\end{equation}
and we assume that the light Higgs doublets
$h_{\rm u}$ (with weak hypercharge $Y=+\Frac{1}{2}$) and
$h_{\rm d}$ (with weak hypercharge $Y=-\Frac{1}{2}$) lie
respectively in $\Phi_{\rm u}$ and $\Phi_{\rm d}$.
As in the~SU(5) case, this superpotential is taken to be valid
already at the Planck scale.
Furthermore, here too it is preferable to work in the basis where
$\mb{\lambda}^{\rm u}$, which is responsible of the
$Q=\Frac{2}{3}$ quark masses, is diagonal.
In analogy with equations~(\ref{eq:m0A0}), at the Planck scale
we take
$$\mb{m}_\Psi^2=m_0^2\One,\qquad
m_{\Phi_{\rm u}}^2=m_{\Phi_{\rm d}}^2= m_0^2,\qquad
 \mb{A}^{\rm u} = \mb{A}^{\rm d} = A_0\One.$$
After renormalization at the unification scale, we have
\begin{equation}\label{eq:Lsoft10}
-\Lag_{\rm soft}=V_{\rm soft}=
\tilde{\Psi}^\dagger \mb{m}_{\Psi\rm G}^2\tilde{\Psi}+
m_{\Phi_{\rm u}\rm G}^2 |\Phi_{\rm u}|^2+
m_{\Phi_{\rm d}\rm G}^2 |\Phi_{\rm d}|^2+
\tilde{\Psi}^T \mb{A}^{\rm u}_{\rm G}
\mb{\lambda}^{\rm u}_{\rm G}\tilde{\Psi}\Phi_{\rm u}+
\tilde{\Psi}^T \mb{A}^{\rm d}_{\rm G}  \mb{\lambda}^{\rm d}_{\rm G}
\tilde{\Psi}\Phi_{\rm d}
\end{equation}
where
\begin{eqnsystem}{sys:softGUT10}
\mb{m}_{\Psi\rm G}^2&=& \diag(m_{\Psi\rm G}^2,m_{\Psi\rm G}^2,
m_{\Psi\rm G}^2-I_{\rm G})\equiv
m_{\Psi\rm G}^2\One -\I_{\rm G},\\
\mb{A}^{\rm d}_{\rm G} &=&\textstyle
\diag(A_{d\rm G},A_{d\rm G},A_{d\rm G}- \frac{5}{7}I'_{\rm G})
\equiv A_{d\rm G}\One-\frac{5}{7}\I'_{\rm G},\\
\mb{A}^{\rm u}_{\rm G} &=&\textstyle
\diag(A_{u\rm G}-\frac{2}{7}I'_{\rm G},A_{u\rm G}-\frac{2}{7}I'_{\rm G}
,A_{u\rm G}-I'_{\rm G}).
\end{eqnsystem}
When possible we also keep the same notation as in the~SU(5) case,
but of course the relations of the various quantities,
e.g. $I_{\rm G}$ in eq.s~(\ref{sys:softGUT10}a) and~(\ref{sys:softGUT5}a),
to the input parameters at the Planck scale change.
These relations in the~SO(10) case are given in appendix~\ref{RGE5},
as function of the one loop coefficient of the gauge $\beta$-function.
In the text we take $b_{\rm G}=-3$.

\section{The low energy lagrangian in SO(10)}
The further scaling down from $M_{\rm G}$ to the weak scale of the
different parameters gives rise to the low energy lagrangian with
the same form as in eq.~(\ref{eq:lowEnLag}), except that
now also the diagonal squared mass matrix of the left handed sleptons
has a split third eigenvalue
\begin{equation}\label{eq:mL10}
\mb{m}_{L\rm G}^2=m_{L\rm G}^2\One-\I_{\rm G}.
\end{equation}
The masses of the third generation sfermions are all reduced relative
to the ones of the first two generations.
For example the $\tilde{\tau}_R$-mass shows approximately
the same pattern
as in the~SU(5) case for a correspondingly lower value
of $\lambda_{t\rm G}$ by a relative amount
$\lambda^{\rm max}_{t\rm G}({\rm SO}(10))/
 \lambda^{\rm max}_{t\rm G}({\rm SU}(5))\approx 0.87$
(see fig.~\ref{fig:mstauR}).

On the other hand, in the fermion mass terms, the symmetry in flavour
space of the~SO(10) coupling $16_i~16_j~10$ gives rise to a symmetric
lepton (or down) mass matrix, so that, in eq.~(\ref{sys:UVde}),
$\mb{U}^{\rm e}=\mb{V}^{\rm e}$.

As before, to calculate the amplitudes for the~LFV processes, it is
convenient to go to the mass-eigenstate basis for the
charged leptons.
At variance with the~SU(5) case, however, this time the term
$\I_{\rm G}$ in~(\ref{eq:mL10}) prevents a counter-rotation also in the
left-handed sleptons.
As a consequence the flavour changing matrix $\mb{V}^{\rm e}$
appears in all the gaugino couplings, which acquire the form
(for all terms involving the charged leptons)
\begin{equation}\label{eq:Lg10}\begin{array}{l}
\Lag_g=\sqrt{2}g'\sum_{n=1}^4\Big[-{1\over2}
\overline{e_L}\,\mb{V}^{\rm e\dagger}\tilde{e}_L  N_n (H_{n\tilde{B}}+
\cot\theta_{\rm W}H_{n\tilde{W}_3})+
\overline{e^c_L}\,\mb{V}^{\rm e\dagger}\tilde{e}_R N_n H_{n\tilde{B}}+
{\rm h.c.}\Big]\\[2mm]
\phantom{\Lag_g=\sqrt{2}g'\sum_{n=1}^4\Big[}
\makebox[0cm][r]{$+g\sum_{c=1}^2\Big[$}
\overline{e_L} \mb{V}^{\rm e\dagger}\tilde{\nu}_L
(\chi_c K_{c\tilde{W}})+{\rm h.c.}\Big]
\end{array}\end{equation}
where $\chi_c$ are the two chargino mass eigenstates, related to
the charged wino by
\begin{equation}\textstyle
\tilde{W} = \sum_{c=1}^2 \chi_c K_{c\tilde{W}},
\end{equation}
and $\tilde{\nu}_L$ is the 3-vector of the left-handed sneutrinos,
which, apart from~${\rm SU}(2)\otimes{\rm U}(1)$ breaking,
are degenerate with the charged left-handed sleptons.

Finally, as in SU(5), there is still a non diagonal `chirality breaking'
scalar mass term
\begin{equation}\label{eq:Lm10}
-\Lag_m^{\rm n.d.}=
(A_e + \mu\tan\beta)
\tilde{e}_R^T   \mb{V}^{\rm e*}\mb{M}^{\rm e}
\mb{V}^{\rm e\dagger}\tilde{e}_L- {\textstyle{1\over2}}
\tilde{e}_R^T\{ {\textstyle{5\over7}}\I'_{\rm G},
\mb{V}^{\rm e*}\mb{M}^{\rm e}\mb{V}^{\rm e\dagger}\}\tilde{e}_L+{\rm h.c.}
\end{equation}
All the LFV couplings in the SO(10) case
are summarized in fig.~\ref{fig:Ver10}.

\begin{figure}
\caption{Lepton flavour violating couplings in SO(10).\label{fig:Ver10}}
\caption[Figura Nove]{Diagrams giving rise to the decay
$\mu\to e\gamma$ in SO(10).
The graphs~\ref{fig:GraMu10}a$_L$,b$_L$,c$_L$,c$'_L$ involving
an external right handed muon and an internal neutralino are not displayed.
As in fig.~\ref{fig:GraMu5}b,c,
in fig.~\ref{fig:GraMu10}b,c,c$'$ an external
photon line is understood.\label{fig:GraMu10}}
\caption[mOH]{Isoplots of ${\rm B.R.}(\mu\to e\gamma)$ in SO(10)
for $m_{\tilde{e}_R}=300\,{\rm GeV}$, $\lambda_{t\rm G}=1.25$
and all other parameters as in fig.~\ref{fig:mU100}.\label{fig:mOH}}
\caption[mOL]{Same as in fig.~\ref{fig:mOH} for $\lambda_{t\rm G}=0.86$.
\label{fig:mOL}}
\caption[mu->e]{Isoplots of ${\rm C.R.}(\mu\to e{\rm~in~Ti})$ in SU(5)
for $m_{\tilde{e}_R}=100$ or $300\,{\rm GeV}$, $\lambda_{t\rm G}=1.4$
and $\tan\beta=2$.\label{fig:mu->e}}
\end{figure}

\section{$\mu\to e\gamma$ in supersymmetric SO(10)}
The Feynman diagrams contributing to $\mu\to e\gamma$ in SO(10)
are shown in figure~\ref{fig:GraMu10} (for vanishing electron mass).
The ones in figures~\ref{fig:GraMu10}a$_R$,a$_L$,d,
with the helicity flip in
the external fermion line, give an amplitude proportional to the muon
mass, whereas the diagrams of
figures~\ref{fig:GraMu10}b$_{L,R}$,c$_{L,R}$,c$'_{L,R}$,
with the helicity flip on the internal fermion line, have a dominant
term proportional to the tau mass.
As such, they dominate the decay rate
in all of the physically allowed space of parameters.
In the approximation of only keeping the terms proportional to
$m_\tau$, the left-handed and the right-handed muon have equal decay
amplitudes, from the diagrams~\ref{fig:GraMu10}b$_R$,c$_R$,c$'_R$
and~\ref{fig:GraMu10}b$_L$,c$_L$,c$'_L$ respectively,
which however do not interfere with each other for vanishing electron mass.

{}From the diagram of figure~\ref{fig:GraMu10}b$_R$ one has
\begin{equation}\begin{array}{ll}
F_2^{({\rm b}_R)} =
\displaystyle\frac{\alpha}{4\pi\cos^2\theta_{\rm W}}m_\tau
V_{\tau\mu}^{\rm e} V_{\tau e}^{\rm e}(V_{\tau\tau}^{\rm e*})^2
(A_e+\mu\tan\beta)\times\\[3mm]
\phantom{F_2^{({\rm b}_R)} =}\times
[G_2(m_{\tilde{\tau}_L}^2,m_{\tilde{\tau}_R}^2)-
G_2(m_{\tilde{e}_L}^2,m_{\tilde{\tau}_R}^2)-
G_2(m_{\tilde{\tau}_L}^2,m_{\tilde{e}_R}^2)+
G_2(m_{\tilde{e}_L}^2,m_{\tilde{e}_R}^2)],
\end{array}
\end{equation}
whereas, from the diagram of figure~\ref{fig:GraMu10}c$_R$,c$'_R$ one has
\begin{equation}\begin{array}{ll}
F_2^{({\rm c}_R)}+F_2^{({\rm c}'_R)} =
\displaystyle\frac{\alpha}{4\pi\cos^2\theta_{\rm W}}m_\tau
V_{\tau\mu}^{\rm e} V_{\tau e}^{\rm e}(V_{\tau\tau}^{\rm e*})^2
(-{\textstyle\frac{5}{7}} I'_{\rm G})\times\\[3mm]
\phantom{F_2^{({\rm c}_R)}+F_2^{({\rm c}'_R)} =}\times
[G_2(m_{\tilde{\tau}_L}^2,m_{\tilde{\tau}_R}^2)
-{1\over2}G_2(m_{\tilde{e}_L}^2,m_{\tilde{\tau}_R}^2)
-{1\over2}G_2(m_{\tilde{\tau}_L}^2,m_{\tilde{e}_R}^2)].
\end{array}
\end{equation}
For the decay rate one has
\begin{equation}\label{eq:F210}
\Gamma(\mu\to e\gamma)=\frac{\alpha}{2} m_\mu^3 |F_2|^2,\qquad
F_2=F_2^{({\rm b}_R)}+F_2^{({\rm c}_R)}+F_2^{({\rm c}'_R)}.
\end{equation}
The isoplots of ${\rm B.R.}(\mu\to e\gamma)$ are shown in
figures~\ref{fig:mOH},~\ref{fig:mOL}.

\begin{figure}
\caption{Box diagrams contributing to $\mu\to e$ conversion in SU(5).
\label{fig:GraMuCapt}}
\end{figure}

\section{$\mu\to e$ conversion}
Keeping only the vector coupling to the nucleus ${\cal N}$, the general
amplitude for $\mu\to e$ conversion process can be written as
\begin{equation}\label{eq:MuCaptAmpl}
{\cal A}=ie^2 [\bar{\cal N}\gamma^\mu{\cal N} ]\,
[\bar{u}_e (g_{1R} \gamma_\mu {\cal P}_R
-g_{2R} {i\sigma_{\mu\nu} q^\nu\over m_\mu}{\cal P}_L)u_\mu]
+(R\leftrightarrow L).
\end{equation}
This amplitude gives rise to the coherent conversion rate
\begin{equation}
\Gamma(\mu\to e) = 4\alpha^5{Z_{\rm eff}^4 \over Z}|F(q)|^2
m_\mu^5(|g_{1R}-g_{2R}|^2 + |g_{1L}-g_{2L}|^2).
\end{equation}
where $Z$ is the charge of the nucleus, $Z_{\rm eff}$ is an effective
charge and $F(q)$ the nuclear form factor~\cite{Titanio}.

In our case, the amplitude receives contributions both from
penguin-type (P) and from box (B) diagrams.
More precisely, to leading order in the lepton and light quark masses,
the penguin diagrams contribute both to $g_1$ and $g_2$, unlike
the box diagrams, which only contribute to $g_1$
\begin{equation}
g_1 = g_1^{\rm B} + g_1^{\rm P},\qquad
g_2 = g_2^{\rm P}.
\end{equation}
If we define, in analogy with eq.~(\ref{eq:AmuSU5}),
the general off-shell $\mu\to e\gamma$ amplitude
${\cal A}_\mu(\mu\to e \gamma)$ as
\begin{equation}
{\cal A}_\mu(\mu\to e\gamma)=
-ie\cdot\bar{u}_e [q^2\,F_{1R}\gamma_\mu {\cal P}_R +
i \sigma_{\mu\nu} q^\nu\,  F_{2R} {\cal P}_L] u_\mu+
(R\leftrightarrow L)
\end{equation}
one has
\begin{eqnsystem}{sys:gF}
g_{1L,R}^{\rm P} &=& Z F_{1L,R}.\\
g_{2L,R}^{\rm P} &=& Z F_{2L,R}/m_\mu. \label{eq:g2F2}
\end{eqnsystem}
In the case of the~SO(10) gauge theory, only the magnetic penguin-type
amplitudes $g_{2L,R}^{\rm P}$ have a term proportional to $m_\tau$,
as discussed in section~9.
Furthermore $g_{2L}^{\rm P}=g_{2R}^{\rm P}\equiv g_2^{\rm P}$.
Therefore, apart from terms of relative order $(m_\mu/m_\tau)^2$,
in view of eq.~(\ref{eq:g2F2}),
\begin{equation}
\Gamma(\mu\to e) =
16 \alpha^4 Z_{\rm eff}^4 Z|F(q)|^2 \Gamma(\mu\to e\gamma).
\end{equation}
In the case of~Ti$^{48}_{22}$, for which~\cite{Titanio}
$Z=22$, $Z_{\rm eff}=17.6$, $|F(q)|=0.54$, taking into account
the experimental value for the capture rate
$\Gamma(\mu{\rm~capture~in~Ti})=
(2.590\pm0.012)\cdot 10^6/{\rm sec}$~\cite{capture}
we obtain
\begin{equation}\label{eq:BRCaptDec10}
{\rm C.R.}(\mu\to e{\rm~in~Ti}) \equiv
\frac{\Gamma(\mu\to e{\rm~in~Ti})}{\Gamma(\mu{\rm~capture~in~Ti})}=
0.5\cdot 10^{-2}\,{\rm B.R.}(\mu\to e\gamma)
\end{equation}
or, normalizing both ratios to
the present upper limits~\cite{expMuDec,expMuCapt}
\begin{equation}\label{eq:BRCaptDec10N}
\frac{{\rm C.R.}(\mu\to e{\rm~in~Ti})}{10^{-12}}= 0.25
\frac{{\rm B.R.}(\mu\to e\gamma)}{4.9\cdot 10^{-11}}.
\end{equation}
Using relation~(\ref{eq:BRCaptDec10}),
the contours of figures~\ref{fig:mOH} and~\ref{fig:mOL} can be relabelled
with values of ${\rm C.R.}(\mu\to e{\rm~in~Ti})$.
This relation holds whenever the $g_1$ form factor
contribution can be neglected.
In the~SO(10) model this is
always the case, giving this relation a wide applicability; wider than the
applicability of the results for the rates themselves. For example, this
relation is independent of $\lambda_{t\rm G}$, the form and values of the
supersymmetry breaking parameters and the form of the RGE above $M_{\rm G}$
(even if the theory becomes non-perturbative).

Contrary to the~SO(10) case, in~SU(5), all the form factors in
eq.~(\ref{eq:MuCaptAmpl}) contribute
in principle at the same general level.

In practice, the contribution of the magnetic form factor $g_2^{\rm P}$
is almost always numerically dominant also in the~SU(5) case,
at least as long as the gaugino mass
parameter $M_2$ is not close to zero or the magnetic form factor
has no accidental cancellation, which may occur for $\mu<0$.
Relative to the penguin contribution $g_1^{\rm P}$, this comes about
because the electric form factor has no term proportional to
$A_e$ and $\mu\tan\beta$.
The box contribution has a quite different structure
from the penguin contribution.
{}From the diagrams of figure~\ref{fig:GraMuCapt} one obtains for
the effective Hamiltonian involving the quark fields
\begin{eqnarray}
{\cal H}^{\rm B}(\mu\to e) &=& -i\frac{\alpha^2}{\cos^4\theta_{\rm W}}
 V_{e\tau}^{\rm e}V_{\mu\tau}^{\rm e*}
[\overline{e_R}\,\gamma_\mu \mu_R]\times\\
&&\times\left[
{1\over36}(c_{u_L}\cdot\overline{u_L}\gamma^\mu u_L +
           c_{d_L}\cdot\overline{d_L}\gamma^\mu d_L)-
{4\over 9}(c_{u_R}\cdot\overline{u_R}\gamma^\mu u_R)-
{1\over 9}(c_{d_R}\cdot\overline{d_R}\gamma^\mu d_R)\right]\nonumber
\end{eqnarray}
with
\begin{eqnsystem}{sys:cudLR}
c_{u_L} &=&  \sum_{n,m=1}^4
[B(m_{\tilde{\tau}_R}^2,m_{\tilde{u}_L}^2,M_n, M_m)-
 B(m_{\tilde{e}_R}^2,   m_{\tilde{u}_L}^2,M_n, M_m)]\times\nonumber\\
&&\times
H_{n\tilde{B}} H_{m\tilde{B}}
(H_{n\tilde{B}}+3 H_{n\tilde{W}_3}\cot\theta_{\rm W})
(H_{m\tilde{B}}+3 H_{m\tilde{W}_3}\cot\theta_{\rm W})\\
c_{d_L} &=&  \sum_{n,m=1}^4
[B(m_{\tilde{\tau}_R}^2,m_{\tilde{d}_L}^2,M_n, M_m)-
 B(m_{\tilde{e}_R}^2,m_{\tilde{d}_L}^2,M_n, M_m)]\times\nonumber\\
&&\times
H_{n\tilde{B}}H_{m\tilde{B}}
(H_{n\tilde{B}}-3 H_{n\tilde{W}_3}\cot\theta_{\rm W})
(H_{m\tilde{B}}-3 H_{m\tilde{W}_3}\cot\theta_{\rm W})\\
c_{u_R} &=& \sum_{n,m=1}^4
[B(m_{\tilde{\tau}_R}^2,m_{\tilde{u}_R}^2,M_n, M_m)-
 B(m_{\tilde{e}_R}^2,   m_{\tilde{u}_R}^2,M_n, M_m)]
H_{n\tilde{B}}^2 H_{m\tilde{B}}^2\\
c_{d_R} &=& \sum_{n,m=1}^4
[B(m_{\tilde{\tau}_R}^2,m_{\tilde{d}_R}^2,M_n, M_m)-
 B(m_{\tilde{e}_R}^2,   m_{\tilde{d}_R}^2,M_n, M_m)]
H_{n\tilde{B}}^2 H_{m\tilde{B}}^2
\end{eqnsystem}
\begin{equation}
B(m_1^2,m_2^2,M_1,M_2)\equiv i(4\pi)^2\int\frac{d^4 k}{(2\pi)^4}
\frac{k^2+2 M_1 M_2}{(k^2-m_1^2)(k^2-m_2^2)(k^2-M_1^2)(k^2-M_2^2)}.
\end{equation}
Consequently, by means of
($N$ is the number of neutrons in the nucleus)
\begin{equation}
\begin{array}{l}
\langle{\cal N}|\overline{u_{L,R}}\gamma_\mu u_{L,R}|{\cal N}\rangle\approx
(Z+N/2)\bar{\cal N}\gamma_\mu{\cal N}\\
\langle{\cal N}|\overline{d_{L,R}}\gamma_\mu d_{L,R}|{\cal N}\rangle\approx
(N+Z/2)\bar{\cal N}\gamma_\mu{\cal N}\\
\end{array}
\end{equation}
one has
\begin{equation}
g_{1R}^{\rm B}=\frac{e^2}{(4\pi\cos^2\theta_{\rm W})^2}
V_{e\tau}^{\rm e}V_{\mu\tau}^{\rm e*}
\frac{1}{72}
[Z(32 c_{u_R}-2c_{u_L}+4c_{d_R}-c_{d_L})+
N(16c_{u_R}-c_{u_L}+8c_{d_R}-2c_{d_L})].
\end{equation}
A numerical calculation shows that the contribution of
the box diagrams to the decay rate goes significantly below the
analogous contribution from the penguins as soon as $M_2$
moves away from zero,
due to the rapid increase of the squark masses in the denominator
and, even more so, to an increase of the magnetic contribution.
In figure~\ref{fig:mu->e} we give the rate for $\mu\to e$ conversion
in the SU(5) case and $\tan\beta=2$.
The numerical results for $\tan\beta=10$ are not shown because they
reproduce the relation~(\ref{eq:BRCaptDec10}) to a good approximation
for any value of the others parameters.

\section{$\tau\to\mu\gamma$}
Very similar considerations to those developed in the previous
sections can be made in the case of the $\tau\to\mu\gamma$ decay.
A main point is the relative difference between the~SU(5)
and the~SO(10) case.

In the~SU(5) case, the amplitude for $\tau\to \mu\gamma$
is simply obtained from ${\cal A}_\mu(\mu\to e\gamma)$,
equations~(\ref{eq:AmuSU5}--\ref{sys:F2abc})
with the replacement of the factor
$m_\mu  V_{\tau \mu}^{\rm e*}V_{\tau  e}^{\rm e}$ by
$m_\tau V_{\tau\tau}^{\rm e*}V_{\tau\mu}^{\rm e}$,
up to negligible terms of relative order $m_\mu/m_\tau$
(remember that in~SU(5) $\tilde{\tau}_L$ and $\tilde{\mu}_L$ are
degenerate to a very good accuracy).
Consequently the following relation holds\footnote{This
equation corrects eq.~(21) of ref.~\cite{BH}.}
\begin{equation}\label{eq:BRmu/tau}
\left.\frac{{\rm B.R.}(\tau\to\mu\gamma)}
{{\rm B.R.}(\mu\to e\gamma)}\right|_{\rm SU(5)} =\!\!
\left|{V_{\tau\tau}^{\rm e} \over V_{e\tau}^{\rm e}}\right|^2
{\rm B.R.}(\tau\to\mu\nu\bar{\nu})
\approx 3\cdot 10^{3}\left({0.77\over y}\right)^{\!\!2}
\left|\frac{0.01}{V_{td}}\right|^2
\end{equation}
with $y$ given in~(\ref{eq:y}).
For given values of the mixing angles, this relation establishes
the relative merit of the searches for the two decay processes
as a possible signal of lepton flavour violation.
With $|V_{td}|=0.01$, the present limit on $\mu\to e\gamma$
(${\rm B.R.}<4.9\cdot 10^{-11}$)~\cite{expMuDec}
is about~30 times better than
the present bound on $\tau\to\mu\gamma$
(${\rm B.R.}<4.2\cdot 10^{-6}$)~\cite{expTauDec}.
Using the relation~(\ref{eq:BRmu/tau}),
the contours of figures~\ref{fig:mU100} and~\ref{fig:mU300} can be
relabelled with values of ${\rm B.R.}(\tau \to \mu \gamma)$.

In~SO(10), the $\mu\to e\gamma$ amplitude is proportional to $m_\tau$ and
is therefore enhanced,
as discussed in section~9.
Consequently the ratio of the branching ratios will be
further suppressed in~SO(10), relative to eq.~(\ref{eq:BRmu/tau}),
by an approximate factor of order $(m_\mu/m_\tau)^2$.

\section{Electric dipole moment of the electron}
It has been pointed out by Dimopoulos and one of us (L.H.)~\cite{DH} that,
in an~SO(10) unified theory, the low energy Lagrangian gives rise to an
electric dipole moment for the neutron and the electron originating from
the phases of the Yukawa couplings.
We concentrate here on the dipole moment, $d_e$, of the electron,
since in this case a very simple relation exists between $d_e$ and
the $\mu\to e\gamma$ rate.
It is clear however that the search for a dipole moment of the neutron
constitutes an independent and equally important signature for the
general effect discussed in this paper.

The full set of diagrams that contribute to the electric dipole
or magnetic moments of the electron coincides with the one
shown in fig.~\ref{fig:GraMu10} with $\mu_L(\mu^c_L)$
replaced by $e_L(e^c_L)$.
In particular, as readily seen from the different dependence
on the~CKM matrix elements, only the diagrams of
figures~\ref{fig:GraMu10}b,c,c$'$
contribute to the electric dipole moment
(with $V_{\tau\mu}^{\rm e}$ replaced by $V_{\tau e}^{\rm e}$),
since they are the only ones with an imaginary part.
These are, on the other hand, the same diagrams that dominate
the $\mu\to e\gamma$ amplitude through their $m_\tau$ dependent
contribution.
As a consequence, the following approximate relation holds between
the form factor $F_2$ defined in eq.~(\ref{eq:F210}) and
the electron dipole moment, $d_e$
\begin{equation}
|d_e|=
e|F_2|\left|{V_{\tau e}^{\rm e}\over V_{\tau\mu}^{\rm e}}\right|
\sin\varphi=
e|F_2|\left|{V_{td}\over V_{ts}}\right|\sin\varphi
\end{equation}
with the~CP violating phase $\varphi$ defined by
$${\rm Im}\,[m_\tau (V_{\tau e}^{\rm e})^2(V_{\tau\tau}^{\rm e*})^2]\equiv
|m_\tau (V_{\tau e}^{\rm e})^2(V_{\tau\tau}^{\rm e*})^2|\sin\varphi.$$
In~SO(10) the electric dipole moment is therefore approximately related
to the $\mu\to e\gamma$ branching ratio by
\begin{equation}\label{eq:de/MuDec}
{|d_e|\over 10^{-27}\,e\cdot{\rm cm}}=1.3 \sin\varphi
\sqrt{\frac{{\rm B.R.}(\mu\to e\gamma)}{10^{-12}}}.
\end{equation}
Using this relation, the contours of
figures~\ref{fig:mOH} and~\ref{fig:mOL} can be relabelled with
values of $|d_e|/\sin\varphi$.
It is interesting to notice that the present upper bound on
$\mu\to e\gamma$ (${\rm B.R.}<4.9\cdot10^{-11}$)
and $d_e$ ($|d_e|<4.3\cdot 10^{-27}\,e\cdot{\rm cm}$)~\cite{expde}
are almost exactly equivalent for $\sin\varphi=\Frac{1}{2}$.

As in~SO(10), in~SU(5) too, the diagrams that could contribute to the
electric dipole moment of the electron are obtained from those of
figure~\ref{fig:GraMu5} by replacing the muon with the electron in the
external line.
This time, however, no electric dipole moment arises since the~CP
violating phase disappears from the product
$V_{\tau e}^{\rm e*}V_{\tau e}^{\rm e}$
of the relevant~CKM matrix elements.

\section{Conclusions}
In this paper we have discussed the lepton flavour violating processes
and electric dipole moments
induced in a supersymmetric unified theory by the large
top Yukawa coupling.
Under the stated assumptions, the experimental study
of these processes provides a very significant test of
supersymmetric unification.
Already the present experimental limits give, especially in the~SO(10)
case, significant restrictions on the allowed
parameter space, often considerably stronger than those inferred from
direct searches of supersymmetric particles.

The main results of this paper are the contour plots for
${\rm B.R.}(\mu \to e\gamma)$ shown in
figures~\ref{fig:mU100} and~\ref{fig:mU300}, for SU(5), and in
figures~\ref{fig:mOH} and~\ref{fig:mOL}, for~SO(10).
The figures~\ref{fig:mU100} and~\ref{fig:mU300}
can also be used for $\tau\to\mu\gamma$ by a
relabelling of the contours using equation~(\ref{eq:BRmu/tau}).
Similarly the contours of
figures~\ref{fig:mOH} and~\ref{fig:mOL}
can be relabelled using~(\ref{eq:BRCaptDec10}) and~(\ref{eq:de/MuDec})
so that they apply to $\mu\to e$ conversion and to $d_e$ respectively.
The case of $\mu\to e$ conversion in SU(5) is shown
in figure~\ref{fig:mu->e}.
These plots cover the entire physical ranges of the parameters $A_e$ and
$M_2$, and show the behaviour for both signs of $\mu$ and for both large
and small values of $\tan \beta$. For large $m_{\tilde{e}_R}$ the~B.R.
decreases as $1/m_{\tilde{e}_R}^4$; however values of $m_{\tilde{e}_R}$
above 400~GeV require a significant amount of fine tuning~\cite{BG}.

\begin{figure}
\caption[mOPl]{Same as in fig.~\ref{fig:mOH} except for the scale of the
initial condition on the RGEs taken at $2.0\cdot 10^{17}\,{\rm GeV}$.
\label{fig:mOPl}}
\end{figure}
\begin{table}
$$\begin{array}{l|ccc}
{\rm GUT}  &\phantom{\ll}\tau\to\mu\gamma &\mu\to e&d_e
\\ \hline
{\rm SU(5)} &\phantom{\ll\,\,}0.03 &
\,\raisebox{-.4ex}{\rlap{$\sim$}}\raisebox{.4ex}{$>$}\, 0.2 &0\\
{\rm SO(10)}&\ll 0.03&\phantom{
\,\raisebox{-.4ex}{\rlap{$\sim$}}\raisebox{.4ex}{$>$}\,}0.2 &2 \sin\varphi
\end{array}$$
\caption[Merits]{Relative merits of various observables
relative to $\mu\to e\gamma$.
All branching ratios, as well as the value of $d_e$, are normalized
to the present limits
(${\rm B.R.}(\mu\to e\gamma) <4.9\cdot 10^{-11}$~\cite{expMuDec},
 ${\rm C.R.}(\mu\to e{\rm~in~Ti})<10^{-12}$~\cite{expMuCapt},
 ${\rm B.R.}(\tau\to\mu\gamma) <4.2\cdot 10^{-6}$~\cite{expTauDec},
 $|d_e|<4.3\cdot 10^{-27}\,e\cdot{\rm cm}$~\cite{expde}).
\label{tab:merits}}
\end{table}

The sensitivities of the various processes to the SU(5) and SO(10) theories
is summarized in table~\ref{tab:merits}, relative to that of
$\mu \to e \gamma$, with all observables
normalized to the present experimental bounds.
For $\sin\varphi<\Frac{1}{2}$ all entries of this table are less
than unity, showing that, for this case, $\mu \to e \gamma$ is
presently the most powerful probe in all cases.
For $\sin \varphi>\Frac{1}{2}$ the electron electric dipole moment
provides the best probe of the SO(10) theory.
The decay $\tau \to \mu \gamma$ will only become competitive with
the construction of a $\tau$ factory.
Future technologies and experimental possibilities should allow an
interesting competition to develop amongst the other three processes.

An additional prediction of this work is that the mass of $\tilde{\tau}_R$ is
suppressed significantly beneath that of  $\tilde{\mu}_R$ and
$\tilde{e}_R$, as can be seen in figure~\ref{fig:mstauR}.
In addition, in~SO(10) the mass of $\tilde{\tau}_L$ is suppressed beneath
that of $\tilde{\mu}_L$ and $\tilde{e}_L$. This result is important for
superpartner searches at $e^+ \; e^-$ colliders: the lightest charged
scalar superpartner is almost certainly a scalar tau.

The sizes of these effects depend on the following two main assumptions:
\begin{itemize}
\item[i)] The value of the top Yukawa coupling at the unification
scale is large;
\item[ii)] The field theoretic renormalization group equations, valid
at the unification scale, can be extrapolated without substantial
modifications up to the reduced Planck mass,
$M_{\rm Pl} = 2.4\cdot10^{17}\,{\rm GeV}$, or to a scale
close to it, e.g.\ the compactification scale of string theory.
\end{itemize}
A value of the top Yukawa coupling at the unification scale less than
one leads to a substantial reduction of the rates, as shown, e.g., in
fig.~\ref{fig:mOL} as compared to fig.~\ref{fig:mOH}.
Also significant, although relatively less important, is the
lowering of the scale for the universal conditions
on the supersymmetry breaking parameters, as exemplified in
figure~\ref{fig:mOPl} where such scale is taken at
$2.0\cdot10^{17}\,{\rm GeV}$.

What other features of the unified model influence our results?
Other than on the gauge group itself, the lepton flavour violation effects
certainly depend on the specific form of the flavour interactions.
In this paper we have studied the two simplest unified sectors that
we know, one in~SU(5) and the other in~SO(10).
Each has the minimal number of flavour Yukawa matrices,
$\mb{\lambda}^{\rm u}$ giving mass to up quarks and
$\mb{\lambda}^{\rm d}$ to down quarks and charged leptons.
It is well known that these Yukawa couplings lead to unrealistic
mass relations between the light fermions.
However, as already discussed in ref.~\cite{BH}, we do not expect
that the necessary modifications of these couplings may lead to significant
suppressions of the lepton flavour violation processes.
They will rather give rise to an increased range of predictions about the
central values discussed here.

The traditional probes of supersymmetric unified theories are provided by
proton decay, neutrino masses and by predictions for quark and charged
lepton masses and mixings. These probes also have model dependences which
arise from the choice of the gauge group and the flavour interactions, as
discussed above. However, for each of these three probes, there is also a
much greater uncertainty than in the lepton flavour violation processes. A
generic unified model has a free parameter for each of the flavour masses
and mixing parameters of the standard model, and hence does not make
predictions for the quark and charged lepton masses. Such predictions only
arise when the form of the flavour interactions are restricted by further
assumptions.

While proton decay and neutrino masses are generally to be expected in
superunified models, the sizes of these signatures are extremely model
dependent.
Consider first the case of proton decay. All superunified models contain
baryon and lepton number violating interactions which couple the quarks
and leptons to a set of superheavy coloured states $H$. The amplitude for
proton decay depends on the mass matrix for these $H$ states.
This is perhaps the least understood, and most model dependent,
feature of superunified theories, because it is directly related to the
problem of why the Higgs doublets are much lighter than $M_{\rm G}$.
Only in one particular model~\cite{DG}, where the Higgs are made light by
an extreme fine tune, has it been possible to relate the $H$ mass to known
parameters of the theory and hence make predictions for the proton decay
rate. In fact the resulting rate is large, and this model is close to
being excluded.
In many other models the matrix structure of the masses for the $H$ states
leads to a large suppression of the proton decay amplitude, which then
becomes gauge dominated, yielding a rate which is expected to be about
four orders of magnitude below present experimental limits.

The three neutrinos frequently acquire small masses in superunified models,
particularly if the gauge group contains~SO(10).
However, the size of these masses is inversely proportional to $\mb{M}_R$,
the Majorana mass matrix for the right-handed neutrinos, which breaks
lepton number and is typically not directly related to known parameters of
the theory. A simple expectation of $M_R \approx  M_{\rm G} $ gives masses
for $\nu_e$, $\nu_\mu$, $\nu_\tau$ which are too small to see in
accelerator or reactor experiments.

By comparison with these great uncertainties, which afflict the traditional
signatures for superunified models, the model dependence of the rates for
$L_e$, $L_\mu$, $L_\tau$
and~CP violating processes discussed here seems quite mild.

We therefore conclude that searches for the $L_i$ and CP violating
signatures discussed in this paper provide the most powerful known probes
of supersymmetric quark-lepton unification with supersymmetry breaking
generated at the Planck scale.
For example, an experiment with a sensitivity of $10^{-13}$ to
${\rm B.R.}(\mu\to e\gamma)$ would probe
(apart from a small region of parameter space where cancellations in
the amplitude occur) the SU(5) model to $\lambda_{t\rm G} = 1.4$ and
$m_{\tilde{e}_R}= 100\,{\rm GeV}$,
and would explore a significant portion of parameters space for
$m_{\tilde{e}_R}=300\,{\rm GeV}$.
In the~SO(10) case, where the present bound on $\mu\to e\gamma$ is already
more stringent than the limits from high energy accelerator experiments, a
sensitivity of $10^{-13}$ would probe the theory to
$\lambda_{t\rm G} = 1.25$ and $m_{\tilde{e}_R}$ close to $1\,$TeV.

\appendix

\begin{table}
$${\rm SU(5)}\phantom{0}\qquad\begin{array}{llr}
b^{\rm u}_g=\{3,3,9\}\phantom{0} & c^{\rm u}=96/5&c^H=24/5 \\
b^{\rm d}_g=\{0,0,3\} & c^{\rm d}=84/5&c^T=36/5
\end{array}$$
$${\rm SO(10)}\qquad\begin{array}{lll}
b^{\rm u}_g=\{4,4,14\} & c^{\rm u}=63/2&c^\Phi=9\phantom{0/0} \\
b^{\rm d}_g=\{0,0,10\} & c^{\rm d}=63/2&c^\Psi=45/4
\end{array}$$
\caption{Values of the RGE coefficients in SU(5) and in SO(10).
\label{tab:c5e10}}
\end{table}

\section{Renormalization from the Planck to the GUT scale}\label{RGE5}
Neglecting all couplings except the gauge and the top Yukawa ones,
the solutions to all the one loop RGEs between
$E_{\rm max}=M_{\rm Pl}$ and
$E_{\rm min}=M_{\rm G}$ can be given analytically.

The RGEs for the dimensionless couplings and for the
dimension-one soft terms are
\begin{eqnsystem}{sys:RGE5}
{d \over {dt}}{1\over\alpha_{5}} &=& 4\pi b_{\rm G}\qquad
{d \over {dt}}{M_5\over \alpha_5}=0   \\
{d \over {dt}}\lambda_t^2 &=&
\lambda_t^2(c^{\rm u} g_5^2-b_t\lambda_t^2)\\
{d \over {dt}}A^{\rm u}_g &=&
c^{\rm u} g_5^2 M_5- b^{\rm u}_g\lambda_t^2 A_t\\
{d \over {dt}}A^{\rm d}_g &=&
c^{\rm d} g_5^2 M_5- b^{\rm d}_g\lambda_t^2 A_t
\end{eqnsystem}
where $t(E)=(4\pi)^{-2}\ln M_{\rm Pl}^2/E^2$,
$g_5$ is the coupling constant of the unification group,
$\alpha_5=g_5^2/4\pi $,
$M_5$ is the gaugino mass, $g=1,2,3$ is the generation number
and the values of the numerical coefficients in~SU(5) and in~SO(10)
are given in table~\ref{tab:c5e10}.
The subscript `5' stands for `unified' rather than for~SU(5).
We also set $b^{\rm u}_3\equiv b_t$.

The full analytic solutions of these equations with
the boundary conditions
$$\alpha_5(M_{\rm G})=\alpha_{\rm G},\qquad
M_5(M_{\rm G}) = M_{5\rm G},\qquad
\lambda_t(M_{\rm G})=\lambda_{t\rm G}$$
and universal $A$-terms at the Planck scale, $A_0$, are
\begin{eqnsystem}{sys:Sol1GUT}
\alpha_5(E) &=&  f_5^{-1}(E)\cdot\alpha_{\rm G}\\
M_5(E) &=& f_5^{-1}(E)\cdot M_{5\rm G}\\
\lambda_t^2(E) &=& \frac{\lambda^{\rm 2max}_t(E)}
{1+\lambda^{\rm 2max}_t(E)
(\lambda_{t\rm G}^{-2}-\lambda^{-2\rm max}_{t\rm G})
f_5^{-c^{\rm u}/b_{\rm G}}(E)}\\
A^{\rm u}_g(E) &=& A_0+x_1^{\rm u}(E) M_{5\rm G} - b^{\rm u}_g I'(E)/b_t\\
A^{\rm d}_g(E) &=& A_0+x_1^{\rm d}(E) M_{5\rm G} - b^{\rm d}_g I'(E)/b_t
\end{eqnsystem}
where the functions $f_5(E)$, $x_n^R(E)$, $\lambda^{\rm max}_t(E)$,
$I'(E)$ are
explicitly defined below in eq.~(\ref{sys:Panta5e10}).

Assuming universal values at the Planck scale
for the dimension-two supersymmetry breaking soft terms,
$m_R^2=m_0^2$,
the one loop RGEs in SU(5)
\begin{eqnarray*}
{d \over {dt}}m_{\bar{F}_g}^2 &=& 2 c^H g_5^2 M_5^2\\
{d \over {dt}}m_H^2 &=&2 c^H g_5^2 M_5^2-
{b_t\over3}\lambda _t^2(2m_{T_3}^2+m_H^2+A_t^2)\\
{d \over {dt}}m_{T_g}^2 &=& 2c^{T} g_5^2 M_5^2-
{b_t\over3}\lambda _t^2(2m_{T_3}^2+m_H^2+A_t^2)\\
\noalign{\hbox{are solved by}}\\[-3mm]
m_{\bar{F}_g}^2(E) &=& m_0^2+x_2^H(E) M_{5\rm G}^2\\
m_H^2(E) &=& m_0^2+x_2^H(E) M_{5\rm G}^2-I(E)\\
m_{T_g}^2(E) &=& m_0^2+x_2^T(E) M_{5\rm G}^2- I(E)\delta_{g3}
\end{eqnarray*}
and in SO(10)
\begin{eqnarray*}
{d \over {dt}}m_\Phi^2 &=& 2c^\Phi g_{5}^2 M_{5}^2-
{4\over14}b_t\lambda _t^2(2m_{\Psi_3}^2+m_\Phi^2+A_t^2)\\
{d \over {dt}}m_{\Psi_g}^2 &=& 2c^\Psi g_{5}^2 M_{5}^2-
{5\over14}b_t\lambda _t^2(2m_{\Psi_3}^2+m_\Phi^2+A_t^2)\\
\noalign{\hbox{by}}\\[-3mm]
m_\Phi^2(E) &=& m_0^2+x_2^\Phi(E) M_{5\rm G}^2-3{4\over14}I(E)\\
m_{\Psi_g}^2(E) &=& m_0^2+x_2^\Psi(E) M_{5\rm G}^2-3{5\over14}I(E)
\delta_{g3}
\end{eqnarray*}
In both cases, we have defined
\begin{eqnsystem}{sys:Panta5e10}
f_5(E)&\equiv& 1+g_{\rm G}^2 b_{\rm G}[t(E)-t(M_{\rm G})]\\
x_n^R(E) &\equiv & \frac{c^R}{b_{\rm G}} [f_5^{-n}(E_{\rm max})-f_5^{-n}(E)]\\
\lambda^{\rm 2max}_t(E) &\equiv &{c^{\rm u}+b_{\rm G}\over b_t}
\frac{g_5^2(E)}{1-
[f_5(E_{\rm max})/f_5(E)]^{1+c^{\rm u}/b_{\rm G}}}\\
I(E)  &\equiv& \rho\Big[m_0^2+{1\over3}(1-\rho)  A_0^2-
{2\over3} (1-\rho)(1-b_t\lambda^{\rm 2max}_t t)A_0 M_{5\rm Pl} \\
&&-{1\over3}[\rho (1-b_t\lambda^{\rm 2max}_t t)^2-b_t c^{\rm u}
\lambda^{\rm 2max}_t g_{5\rm G}^2 t^2] M_{5\rm Pl}^2\Big]\nonumber\\
I'(E) &\equiv&\rho[A_0-(1-b_t\lambda^{\rm 2max}_t t) M_{5\rm Pl}]
\end{eqnsystem}
and  $\rho(E)\equiv\lambda_t^2(E)/\lambda^{\rm 2max}_t(E)<1$.
{}From eq.s~(\ref{sys:Panta5e10}), one learns that the main factor
that determines the size of the lepton flavour breaking
parameters $I$ and $I'$ is the overall factor $\rho$.
In turn, $\rho$ is only weakly dependent on the $\beta$-function
coefficient $b_{\rm G}$ (see figure~\ref{fig:ltmaxbG}).

For the numerical values at $M_{\rm G}$
of the different quantities defined above
we take
$$\alpha_{\rm G}=1/24,\qquad
M_{\rm G}=2.0\cdot 10^{16}\,{\rm GeV},\qquad
M_{\rm Pl}=2.4\cdot 10^{18}\,{\rm GeV},\qquad b_{\rm G}=-3,$$
so that $t_{\rm G}=0.0606$
(a subscript `G' on the  various functions of $E$
indicates that they are evaluated at $M_{\rm G}$).
In~SU(5) $\lambda^{\rm max}_{t\rm G}=1.56$ and
\begin{eqnsystem}{sys:NPanta5}
I _{\rm G}\equiv I (M_{\rm G}) &=&
\rho_{\rm G}\Big[m_0^2+{1\over3}(1-\rho_{\rm G})  A_0^2+
0.198(1-\rho_{\rm G}) A_0 M_{5\rm G} +
(0.224-0.029\rho_{\rm G})M_{5\rm G}^2\Big]\\
I'_{\rm G}\equiv I'(M_{\rm G}) &=&\rho_{\rm G}[A_0+0.298 M_{5\rm G}]
\end{eqnsystem}
while in~SO(10) $\lambda^{\rm max}_{t\rm G}=1.36$ and
\begin{eqnsystem}{sys:NPanta10}
I _{\rm G}\equiv I (M_{\rm G}) &=&
\rho_{\rm G}\Big[m_0^2+{1\over3}(1-\rho_{\rm G})  A_0^2+
0.343(1-\rho_{\rm G}) A_0 M_{5\rm G} +
(0.435-0.088\rho_{\rm G})M_{5\rm G}^2\Big]\\
I'_{\rm G}\equiv I'(M_{\rm G}) &=&\rho_{\rm G}[A_0+0.515 M_{5\rm G}].
\end{eqnsystem}

\begin{table}
$$\begin{array}{c|ccccc|ccc||c||ccc}
\phantom{-}b_i&c_i^Q&c_i^u&c_i^d&c_i^L&c_i^e&
c_i^{\rm u}&c_i^{\rm d}&c_i^{\rm e}&i,g&
b^{\rm u}_g&b^{\rm d}_g&b^{\rm e}_g\\[0.5mm] \hline
\phantom{-}{33\over5} &{1\over30}&{8\over15}&{2\over15}&\vphantom{X^{X^X}}
{3\over10}&{6\over5}& {13\over15} &{7\over15}&{9\over5}&1&3&0&0\\
\phantom{-}1&{3\over2}&0&0&{3\over2}&0&3&3&3&2&3&0&0\\
-3&{8\over3}&{8\over3}&{8\over3}&0&0&{16\over3}&{16\over3}&0&3&6&1&0
\end{array}$$
\caption{Values of the RGE coefficients in the MSSM.\label{Tab:bcude}}
\end{table}

\section{Renormalization in the MSSM}\label{RGEMSSM}
Neglecting all couplings except the gauge and the top Yukawa ones,
the solutions to all the one loop RGEs between
$E_{\rm max}=M_{\rm G}$ and
$E_{\rm min}=M_Z$ may be written in terms of analytic
functions and only one
function, $\lambda^{\rm max}_t(E)$,
calculable only numerically~\cite{RGEinMSSM}.

The RGEs for the dimensionless couplings are
\begin{eqnsystem}{sys:MSSMeq0}
\frac{d}{dt}\frac{1}{\alpha_i} &=& 4\pi b_i\\
\frac{d}{dt}\lambda_t^2 &=&\lambda_t^2(c_i^{\rm u} g_i^2-b_t\lambda_t^2)
\end{eqnsystem}
where $i=1,2,3$ runs over the three factor
in the Standard Model gauge group
${\rm U}(1)\otimes\SU(2)\otimes\SU(3)$,
$t(E)=(4\pi)^{-2}\ln M_{\rm G}^2/E^2$ and
the values of the coefficients are shown in table~\ref{Tab:bcude}.

The solutions with boundary conditions
$\alpha_i(M_{\rm G})=\alpha_{\rm G}$ and
$\lambda_t(M_{\rm G})=\lambda_{t\rm G}$ are
\begin{eqnsystem}{sys:MSSMsol0}
\alpha_i(E) &=& f_i^{-1}(E)\cdot \alpha_{\rm G} \\
\lambda_t^2(E) &=& \frac{\lambda^{\rm 2max}_t(E)}{1+
\lambda^{\rm 2max}_t(E)/\lambda_{t\rm G}^2 E_{\rm u}(E)}
\label{eq:lambda_t}
\end{eqnsystem}
where $f_i(E)\equiv1+b_i g_{\rm G}^2 t(E)$ and
\begin{equation}\label{eq:EF}
E_\alpha(E)\equiv\prod_i f_i^{c_i^\alpha/b_i}(E),\qquad
F_{\rm u}(E)\equiv 2\int^{\ln M_{\rm G}}_{\ln E}\!\! E_{\rm u}(E) d\ln E,
\qquad
\lambda^{\rm 2max}_t(E) = {E_{\rm u}(E)\over b_t F_{\rm u}(E)}
\end{equation}
The Yukawa couplings of the fermions in the diagonal basis scale as
$\lambda^\alpha_g(E)=
\lambda^\alpha_g(M_{\rm G})\cdot y^{b_g^\alpha} E_\alpha^{1/2}$
where
$g=1,2,3$ is the generation number,
$\alpha={\rm u},\,{\rm d},\,{\rm e}$
and
\begin{equation}\label{eq:y}
y(E)\equiv\exp\bigg[-\int_{\ln E}^{\ln M_{\rm G}}
\frac{\lambda_t^2(E')}{16\pi^2}d\ln E'\bigg]=
[1-\rho(E)]^{1/2b_t},\qquad
\rho(E)\equiv{\lambda_t^2(E)\over \lambda^{\rm 2max}_t(E)}<1.
\end{equation}
The factor $y$ used in the text is given by $y\equiv y(M_Z)$.

The RGEs for the three gaugino masses $M_i$,
the supersymmetric $\mu$-term and the $A$ terms are
\begin{eqnsystem}{sys:MSSM1eq}
\frac{d}{dt}\frac{M_i}{\alpha_i} &=& 0\\
\frac{d}{dt}\mu &=&{\textstyle\frac{1}{2}}
(2c_i^h g_i^2 - b_1^{\rm u} \lambda_t^2)\mu\\
\frac{d}{dt}A_{u,g} &=& c_i^{\rm u} g^2_i M_i-
b^{\rm u}_g\lambda^2_t A_{u,g}\\
\frac{d}{dt}A_{d,g} &=& c_i^{\rm d} g^2_i M_i-
b^{\rm d}_g\lambda^2_t A_{d,g}\\
\frac{d}{dt}A_{e,g} &=& c_i^{\rm e} g^2_i M_i
\end{eqnsystem}
with all the various coefficients listed in table~\ref{Tab:bcude}.
The solutions are
\begin{eqnsystem}{sys:MSSM1sol}
M_i(E) &=& f_i^{-1}(E)\cdot M_{5\rm G}\\
\mu(E) &=& \mu(M_{5\rm G})\cdot y^{b^{\rm u}_1}(E) E_h(E)\\
A^{\rm u}_g(E) &=& A^{\rm u}_{g\rm G} + x_1^{\rm u}(E) M_{5\rm G}-
b^{\rm u}_g I'(E)/b_t\\
A^{\rm d}_g(E) &=& A^{\rm d}_{g\rm G} + x_1^{\rm d}(E) M_{5\rm G}-
b^{\rm d}_g I'(E)/b_t\\
A^{\rm e}_g(E) &=& A^{\rm e}_{g\rm G} + x_1^{\rm e}(E) M_{5\rm G}
\end{eqnsystem}
with $x_1^R(E)$ and $I'(E)$ defined below in~(\ref{sys:PantaMSSM})
and $E_h(E)$ in~(\ref{eq:EF}).

The RGEs for the dimension-two soft parameters of a representation
$R=\{Q,u,d,e,L=h\}$ are
\begin{eqnsystem}{sys:MSSM2eq}
\frac{d}{dt} m_R^2 &=& 2 c_i^R g^2_i M_i^2\\
\noalign{\hbox{except for the multiplets $h_{\rm u}$, $\tilde{Q}_3$ and
$\tilde{t}$ involved in the top Yukawa coupling.
For them}}\nonumber\\[-3mm]
\frac{d}{dt} m_{h_{\rm u}}^2  &=& 2 c_i^{h} g^2_i M_i^2-
{1\over2}b_t\lambda^2_t (A_t^2+3m^2)\\
\frac{d}{dt} m_{\tilde Q_3}^2  &=& 2 c_i^{Q} g^2_i M_i^2-
{1\over6}b_t\lambda^2_t (A_t^2+3m^2)\\
\frac{d}{dt} m_{\tilde t}^2  &=& 2 c_i^{u} g^2_i M_i^2-
{1\over3}b_t\lambda^2_t (A_t^2+3m^2)\\
\noalign{\hbox{where
$m^2(E)\equiv[m_{h_{\rm u}}^2(E)+
m_{\tilde Q_3}^2(E)+ m_{\tilde t}^2(E)]/3$.
The solutions are}}\nonumber \\[-3mm]
m_R^2(E) &=& m_R^2(M_{\rm G})+x_2^R M_{5\rm G}^2\\
\noalign{\hbox{except when
the top Yukawa coupling appears, where}}\nonumber\\[-3mm]
m_{h_{\rm u}}^2(E) &=& m_{h_{\rm u}}^2(M_{\rm G})+
x_2^h(E) M_{5\rm G}^2-{1\over2}3I(E)\\
m_{\tilde Q_3}^2(E) &=& m_{10_3}^2(M_{\rm G})+
x_2^Q(E) M_{5\rm G}^2-{1\over6}3I(E)\\
m_{\tilde t}^2(E) &=& m_{10_3}^2(M_{\rm G})+
x_2^u(E) M_{5\rm G}^2-{1\over3}3I(E)
\end{eqnsystem}
and
\begin{eqnsystem}{sys:PantaMSSM}
x_n^R(E) &\equiv&\sum_{i=1}^3{c_i^R\over b_i}
[f_i^{-n}(E_{\rm max})-f_i^{-n}(E)]\\
I(E) &\equiv& \rho \left[m^2(M_{\rm G})+{1\over 3}(1-\rho) A_{t\rm G}^2-
{2\over3}(1-\rho)(1-b_t\lambda^{\rm 2max}_t t) M_{5\rm G}A_{t\rm G}-
\right.\\
&&\left.{1\over3}[\rho(1-b_t\lambda^{\rm 2max}_t t)^2-
b_t\lambda^{\rm 2max}_t t^2(c_i^{\rm u} g_i^2)]M_{5\rm G}^2\right]
\nonumber\\
I'(E) &\equiv& \rho[A_{t\rm G}-(1-b_t\lambda^{\rm 2max}_t t)M_{5\rm G}]
\end{eqnsystem}
Notice that, apart form obvious replacements, $x_n^R$, $I$ and $I'$
maintain exactly the same form as in eq.~(\ref{sys:Panta5e10}).

The numerical values at $M_Z$
of the different quantities defined above
are $t(M_Z)=t_Z=0.418$
(a subscript `$Z$' on the  various functions of $E$
indicates that they are evaluated at $M_Z$),
$$E_{{\rm u}Z}=13.6,\qquad
b_t F_{{\rm u}Z}=10.5,\qquad \lambda^{\rm max}_{tZ}=1.14$$
\begin{eqnsystem}{sys:NPantaMSSM}
I_Z  &=&  \rho_Z\, m(M_{\rm G}^2)+{\rho_Z\over 3}(1-\rho_Z) A_{t\rm G}^2
+1.50\rho_Z(1-\rho_Z) M_{5\rm G}A_{t\rm G}+
\rho_Z (4.37-1.70\rho_Z)M_{5\rm G}^2\\
I'_Z &=& \rho_Z[A_{t\rm G}+2.26 M_{5\rm G}]
\end{eqnsystem}
$$\begin{array}{lllll}
&f_{1Z}=2.44&f_{2Z}=1.22&f_{3Z}=0.343&\\
&x_{1Z}^{\rm u}=4.02 & x_{1Z}^{\rm d}=3.98&x_{1Z}^{\rm e}=0.700 &\\
x_{2Z}^Q=7.16&x_{2Z}^u=6.73&x_{2Z}^d=6.68&x_{2Z}^e=0.151&
x_{2Z}^L=x_{2Z}^h=0.528
\end{array}$$
Finally the Higgs doublets mass parameters
$m_{\rm d}^2\equiv m_{h_{\rm d}}^2(M_Z)$ and
$m_{\rm u}^2\equiv m_{h_{\rm u}}^2(M_Z)$
defined in eq.~(\ref{eq:mumdmud})
may be expressed in terms of the universal
supersymmetry breaking parameters as
\begin{eqnsystem}{sys:mumd}
&m_{\rm d}^2=m_0^2+(x_{2\rm G}^H + x_{2Z}^h) M_{5\rm G}^2,~\qquad
&m_{\rm u}^2=m_{\rm d}^2-({\textstyle\frac{3}{2}}I_Z+I_{\rm G}),\\
\noalign{\hbox{in~SU(5), while, in~SO(10)}}\nonumber\\[-3mm]
&m_{\rm d}^2=m_0^2+(x_{2\rm G}^\Phi + x_{2Z}^h) M_{5\rm G}^2,\qquad
&m_{\rm u}^2=m_{\rm d}^2-
({\textstyle\frac{3}{2}}I_Z+{\textstyle\frac{6}{7}}I_{\rm G}).
\end{eqnsystem}

\end{document}